\documentclass{aa} 
\usepackage{txfonts}

\usepackage[T1]{fontenc}
\usepackage[english]{babel}
\usepackage[usenames]{xcolor}
\usepackage{mathtools}
\usepackage{url}
\usepackage{soul}

\usepackage{graphicx}	
\usepackage{amsmath}	
\usepackage{amssymb}	
\usepackage[normalem]{ulem}
\usepackage[hyperindex,pdftex]{hyperref}
\usepackage{deluxetable}
\usepackage{chemformula} 

\bibpunct{(}{)}{;}{a}{}{,} 

\hypersetup{linkcolor=red,citecolor=green,filecolor=cyan,urlcolor=magenta}
\usepackage{amstext}
\begin{document}

\title{Effect of mantle oxidation state and escape upon the evolution of Earth's magma ocean atmosphere}

\author{Nisha Katyal\inst{1}
   \and Gianluigi Ortenzi\inst{2,3}
   \and John Lee Grenfell\inst{2}
   \and Lena Noack \inst{3}
   \and Frank Sohl\inst{2}
   \and Mareike Godolt \inst{1,2}
   \and Antonio Garc{\'i}a Mu{\~n}oz \inst{1}
   \and Franz Schreier \inst{4}
   \and Fabian Wunderlich \inst{1,2}
   \and Heike Rauer\inst{1,2,5}}

\institute{Department of Astronomy and Astrophysics, Berlin Institute of Technology, Hardenbergstr. 36, 10623 Berlin, Germany\\ \email{n.katyal@tu-berlin.de}
  \and Institute of Planetary Research, German Aerospace Center, Rutherfordstr. 2, 12489 Berlin, Germany
\and  Department of Earth Sciences, Free University of Berlin, Malteserstr. 74-100, 12249 Berlin, Germany
\and Remote Sensing Technology Institute, German Aerospace Center, 82234 Oberpfaffenhofen, Germany
  \and Department of Planetary Sciences and Remote Sensing, Free University of Berlin, Malteserstr. 74-100, 12249 Berlin, Germany}

\date{Submitted on 2020/06/28 ; Accepted for publication on 2020/09/29}


\abstract
{The magma ocean period was a critical phase determining how Earth's atmosphere developed into habitability. However, there are major uncertainties in the role of key processes such as outgassing from the planetary interior and escape of species to space that play a major role in determining the atmosphere of early Earth. } 
{We investigate the effect of outgassing of various species and escape of \ch{H2} for different mantle redox states upon the composition and evolution of the atmosphere for the magma ocean period.}
{We included an important new atmosphere-interior coupling mechanism: the redox evolution of the mantle, which strongly affects the outgassing of species. We simulated the volatile outgassing and chemical speciation at the surface for various redox states of the mantle by employing a C-H-O based chemical speciation model combined with an interior outgassing model.
We then applied a line-by-line radiative transfer model to study the remote appearance of the planet in terms of the infrared emission and transmission. Finally, we used a parameterized diffusion-limited and XUV energy-driven atmospheric escape model to calculate the loss of \ch{H2} to space.}
{We have simulated the thermal emission and transmission spectra for reduced and oxidized atmospheres present during the magma ocean period of Earth. Reduced/thin atmospheres consisting of \ch{H2} in abundance emit more radiation to space and have a larger effective height than oxidized/thick atmospheres, which are abundant in \ch{H2O} and \ch{CO2}.  We obtain that the outgassing rates of \ch{H2} from the mantle into the atmosphere are a factor of ten times higher than the rates of diffusion-limited escape to space. We estimate the timescale of total mass loss of outgassed \ch{H2} via escape to be few tens
 of million years, which is comparable to other studies.}
{Our work presents useful insight into the development of the terrestrial atmosphere during the magma ocean period and provides input to guide future studies that discuss exoplanetary interior compositions and their possible links with atmospheric compositions that might be estimated from observed infrared spectra by future missions.}

\titlerunning{Effect of mantle oxidation state upon atmospheric evolution}
\authorrunning{Katyal et al.}


\maketitle

\section{Introduction} 
\label{sec:intro}

Understanding how the early atmosphere of Earth emerged into habitable conditions is a central question not only for addressing our own origins, but also for interpreting the fascinating new data in exoplanetary science  \citep[for recent reviews, see, e.g.,][]{Jont2019,Nikku2019}. Magma oceans (MOs) consisting of hot and molten silicates in the mantle exist during the planetary accretion phase  \citep{Elkins2012} and facilitate the formation of an atmosphere by outgassing of volatile species such as \ch{H2O}, \ch{CO2}, \ch{H2}, \ch{CH4} , and \ch{CO} . Thus, the MO period likely represents a key juncture on the pathway to habitability because atmosphere-interior couplings  \citep[e.g., ][]{Elkins2008,Lebrun2013,Hamano2013,Schaefer2016} established during this phase could have set the stage for subsequent atmospheric evolution. 
The duration of the magma ocean phase is potentially affected by the absence or presence of an atmosphere and the interior dynamics of the mantle \citep[e.g.,][]{Lebrun2013,Nasia2019} that affect the thermal spectral evolution of the planet \citep[e.g.,][]{Hamano2015,Katyal2019}. Other uncertainties such as the initial volatile inventory of the mantle, for example, the amount of \ch{H2O} stored in the mantle \citep{Genda2017a,Raymond2019}, and compositional constraints such as the mantle oxygen fugacity \citep{Hirsch2012,Schaefer2017} are also important quantities that might critically affect the oxidation state of outgassed volatiles and hence the subsequent climate and compositional evolution of the atmosphere. 

The budget and origin of volatiles in the mantle of Earth is one of the main questions in geo-science \citep[see, e.g.,][]{Raymond2019}.
Recent studies suggest quite diverse scenarios that range from a wet, late accretion to a dry accretion to which water was added mostly during the ``late veneer", which is proposed to have occurred roughly around 80-130 Myr after the isolation of protoplanetary nebula at around 4.5 Gyr before present \citep{Albarade2013}. 
Abundances of highly siderophile elements (HSE) and other volatile species such as H$_2$O suggest that volatile-rich material was added to Earth after its core formed \citep{Albarede2009}. Measurements of the abundances of water and carbon in the bulk silicate Earth (BSE) show that the late veneer might indeed have supplied 20\% -\ 100\%\ of the budget of the total hydrogen (H) and carbon (C) in the BSE \citep{WangBecker2013}. Some recent studies do not support the late veneer as the main source of water on Earth \citep{Dauphas2017,Kleine2017,Kleine2018}  and argue that water was likely added to Earth already during the waning stages of accretion and thus was already present during the late veneer. Interestingly, the reaction of terrestrial water and Fe from the late veneer produced much hydrogen as a byproduct \citep{Zahnle_impact2020}. There is increasing isotopic evidence indicating that the late veneer might also have supplied a significant amount of reducing material onto the surface of Earth and into the atmosphere \citep{Genda2017a,Greenwood2018,Zahnle_impact2020}.

The impact-degassing studies by \cite{Schaefer2007,Hashimoto2007,Schaefer2010} and \cite{Schaefer2012} also in general favored the outgassing of reduced gases upon impacts by  certain types of reduced meteoritic materials such as ordinary and enstatite chondrites. Furthermore, changes in chondrites delivery rate are likely to affect the redox state of the mantle and hence the atmospheric composition and amount, as suggested recently by \cite{Schaefer2017}. Their work also showed that the variation in oxygen fugacities for mixtures of primitive meteoritic material could lead to atmospheres ranging from highly oxidizing to highly reducing ones \citep[also see][]{Hirsch2012}. 

The redox state of the minerals and melts is related to the oxygen fugacity f\ch{O2} of the system that is equivalent to the partial pressure of the gas or the availability of oxygen \citep{Gaillard2015}. This quantity is poorly constrained for the early Earth \citep{frost2008}.
The current understanding is that before the formation of the Earth core $\sim$4.56 Gyr before present \citep{Stevenson1983}, the lower f\ch{O2} of the upper mantle was approximately three log units below the reducing iron-wüstite (IW) buffer, that is, IW-3  and it evolved to a higher fugacity value that was approximately equivalent to the current mantle of Earth, which resembles the quartz-fayalite-magnetite (QFM) buffer, that is, IW+3.5 \citep{Wood1990,Neill1991}.  Direct
geological evidence regarding the redox state of the mantle  is sparse, and it is not clear when the Earth may have been oxidized during the Hadean \citep{Trail2011}. However, the geological record mainly suggests that the upper mantle was initially reduced and became progressively oxidized between 4.6 and 3.9 Gyr before present \citep{Kasting1993, delano2001redox, Schaefer2017}, with a possible later further increase in redox state by $\approx$ 1.3 at the end of the Archean \citep{sonja2016}. 

Previous interior modeling studies investigating atmospheres at and around the time of the magma ocean \citep[e.g.,][]{Lebrun2013,Majum2017,Nasia2019} have considered outgassing of volatiles such as H$_2$O and CO$_2$, which means that they considered the mantle to be constantly oxidized. These studies suggested outgassed pressures of H$_2$O and CO$_2$ ranging from a few tens to hundreds of bar depending upon uncertainties, for instance, in the initial volatile content, the timing of the MO, and internal properties of the mantle. Recent modeling studies have now started to investigate the effect of the redox state on the outgassing. \citet{Pahlevan2019} investigated the effect of f\ch{O2} on the outgassing and atmosphere losses and provided evidence for an early oxidation of silicate Earth using the D/H of the oceans. Most recently,  \citet{Ortenzi2020} calculated the expected outgassing rates for the reduced and oxidized rocky planets and focused on exploring the observational constraints for the atmospheric (and interior) redox state of exoplanets. 

Hydrodynamic escape of \ch{H2} during this time ($\sim$4.5 Gyr before present) was likely not sufficient to desiccate the planet \citep{Kasting1993}, so that conditions remained wet after the Moon-forming impact and crystallization of the final MO \citep{Lammer2018}. Hydrogen would therefore have been a sufficiently major component of an accretionary steam-based atmosphere during the MO phase. However, hydrogen evolution around the time of the MO is poorly understood because of the uncertainties in accretion, outgassing, ingassing, and escape \citep[see, e.g.,][]{Tian2015Rev}. Several studies have considered the energy-limited hydrodynamic escape of hydrogen from the early atmosphere (Archean) of Earth \citep{Tian2005a,Kuramoto2013,Zahnle2019} and calculated somewhat differing results based on including or excluding the effect of \ch{H2} diffusive flux on the gas-density profile and processes such as radiative cooling and thermal conduction that these escape models treat, for example. \citet{Johnstone2019} have also discussed uncertainties in the energy-limited mass-loss formula when applied to the early atmospheres. For rocky planets around M dwarfs, large amounts of water could be lost depending on the uncertainties in the stellar luminosity, for instance \cite[e.g.,][]{TianIda2015,Schaefer2016}. 

 We apply a coupled suite of interior and atmospheric models to investigate outgassing and 
 escape during the MO period and study the effect of varying the mantle fugacity on the atmospheric evolution. We also calculate theoretical atmospheric spectra that serve as a link for observations with early Earth-sized planets.  In Section~\ref{meth} we present the volatile speciation model and the atmospheric model. Section~\ref{sce} presents the scenarios we adopted. Section~\ref{s3} provides results of the effect of the redox state of the mantle on the infrared emission and transmission spectra. The interplay between outgassing and atmospheric loss of \ch{H2} is also presented in this section. In Section~\ref{s4} we discuss our findings. Finally, we provide a conclusion of our work in Section~\ref{s6}.  


\section{Methods and models} \label{meth}

\subsection{C-H-O based volatile speciation model}\label{s2}

The outgassing and volatile chemical speciation were simulated following the equilibrium and mass balance method \citep{French1966,holloway_1981, Fegley2013,gaillard2014theoretical,Schaefer2017} that has recently been presented in detail by \cite{Ortenzi2020}. We calculated the outgassed composition of the volatiles considering a broad range of temperatures, pressures, and redox states. The four common petrological buffers used in the literature and their mantle oxidation states are 
\begin{equation}
\ch{2 Fe + SiO2 + O2 <=> Fe2SiO4}, \hfill \rm{QIF}
\end{equation}


\begin{equation}
\ch{2 Fe + O2 <=> 2 FeO}, \hfill \rm{IW}
\end{equation}


\begin{equation}
\ch{3 Fe2SiO4 + O2 <=> 2 Fe3O4 + 3 SiO2}, \hfill \rm{QFM}
\end{equation}
\begin{equation}
\ch{Ni + 1/2 O2 <=> NiO}. \hfill \rm{NiNiO}
\end{equation}

Quartz-iron-fayalite (QIF) and IW buffers represent reduced conditions, and quartz-fayalite-magnetite (QFM) and nickel-nickel-oxide (NiNiO) describe the oxidizing redox states. The oxidation state (oxygen fugacity) for these buffers was calculated following the parameterization from \cite{Holloway1992},
\begin{equation}
\label{ox_fug_holloway_92}
\log_{10}\rm f\ch{O2}= A - B/T + C(P-1)/T \; + Z,
\end{equation}
where the pressure (P) is in bars, the temperature (T) in is Kelvin, and the parameters  A, B and C are defined in Table \ref{table_fo2_holloway_92_parameters}.  Z is a positive or negative number denoting the deviation from the fugacity values with respect to the buffers, as stated in the table. We mainly used the IW buffer and chose Z to be -4 (highly reducing), 0 (reducing), and 4 (oxidizing) in order to investigate the effect of a range of redox mantle states on outgassing.

\begin{table}
\centering
    \caption{Data collected from \cite{Holloway1992} to obtain the oxygen fugacity of the mantle buffers as listed.  
    \label{table_fo2_holloway_92_parameters}}
\begin{tabular}{c c c c}
\hline\hline
    {Buffer} & {A} & {B} &
    {C}\\
\hline    
 QIF\tablefootmark{a} & 7.679 & 29673 & 0.05 \\ 
 IW\tablefootmark{b} & 6.899 & 27714 & 0.05   \\
 QFM\tablefootmark{c} & 8.555 & 24014 & 0.092   \\
 NiNiO\tablefootmark{d} & 8.951 & 24556 & 0.046  \\
\hline
\end{tabular}
   \tablefoottext{a}{Quartz-iron-fayalite}
    \tablefoottext{b}{iron-w{\"u}stite}
    \tablefoottext{c}{quartz-fayalite-magnetite}
    \tablefoottext{d}{nickel-nickel-oxide}
\end{table}

By simulating the oxidation state, that is, the oxygen fugacity  f\ch{O2} of the system, we are able to simulate the gas chemical speciation via the following equilibria:

\begin{equation}\label{eq:C}
    \ch{CO + 1/2 O2 <=> CO2}, \; \rm and
\end{equation}

\begin{equation}\label{eq:H}
\ch{2 H2 + O2 <=> 2 H2O}.
\end{equation}




To calculate the ratio between the carbon species (Eq.~\ref{eq:C}), we considered the equilibrium constant $K_1$ for the equilibrium as
\begin{equation}
     K_1 = \exp\left(\frac{-\Delta_rG^0_1}{RT}\right) = \frac{X_{\ch{CO2}}}{X_{\ch{CO}}}\frac{1}{\rm f\ch{O2}^{1/2}},
\end{equation}
where $R$ is the universal gas constant (8.314 J K$^{-1}$ mol$^{-1}$), $T$ is the temperature of the outgassed material in Kelvin, and $\Delta_rG^0_1$ is the Gibbs free energy of the reaction in Eq. (\ref{eq:C}), where
\begin{equation}
    \Delta_rG^0_1 = \Delta_fG^0_{\ch{CO2}} - \Delta_fG^0_{\ch{CO}}.
\end{equation}

Substituting $\Delta_rG^0_1$ and the calculated f\ch{O2} in Eq. \eqref{eq:C}
gives us all the necessary parameters needed to calculate the carbon species. 
For Eq. \eqref{eq:H}, the Gibbs free energy of reaction $\Delta_rG^0_2$ is related only to the Gibbs free energy of formation of water ($\Delta_fG^0_{\ch{H2O}}$), 

\begin{equation}
\Delta_rG^0_2 = 2\Delta_fG^0_{\ch{H2O}}.
\end{equation}


The values of $\Delta_fG^0$ for the different species were
calculated following the example of \cite{Fegley2013} and compared to the literature \citep{janaf}. 

 Similarly, the abundances of \ch{H2} and \ch{H2O} Eq. \eqref{eq:H} are related to the equilibrium constant $K_2$ and the fugacity f\ch{O2} via

\begin{equation}
\left(\frac{X_{\ch{H2O}}}{X_{\ch{H2}}}\right)^2 = K_2 \; \rm f\ch{O2}, 
\label{outgassing}
\end{equation}
where $X_{\ch{H2O}}$ and 
${X_{\ch{H2}}}$ are the mole fractions of \ch{H2O} and \ch{H2} and are related to the partial pressure of each of the species. The rate of outgassing of \ch{H2O} can therefore be related to the rate of outgassing of \ch{H2} as
\begin{equation}
    r_{\ch{H2O}} = R_1 \,\ r_{\ch{H2}},
    \label{rH2}
\end{equation}
where $R_1=(K_2 \; \rm f\ch{O2})^{0.5}$, $r_{\ch{H2}}$ , and $r_{\ch{H2O}}$ are the outgassing rates of \ch{H2} and \ch{H2O} in units of m$^{-2}$ s$^{-1}$, respectively.

\subsection{Coupled interior-atmospheric evolution}\label{pres}

\begin{figure}[!hbt]
  \includegraphics[trim=0 12cm 0 3cm ,clip,width=\hsize]{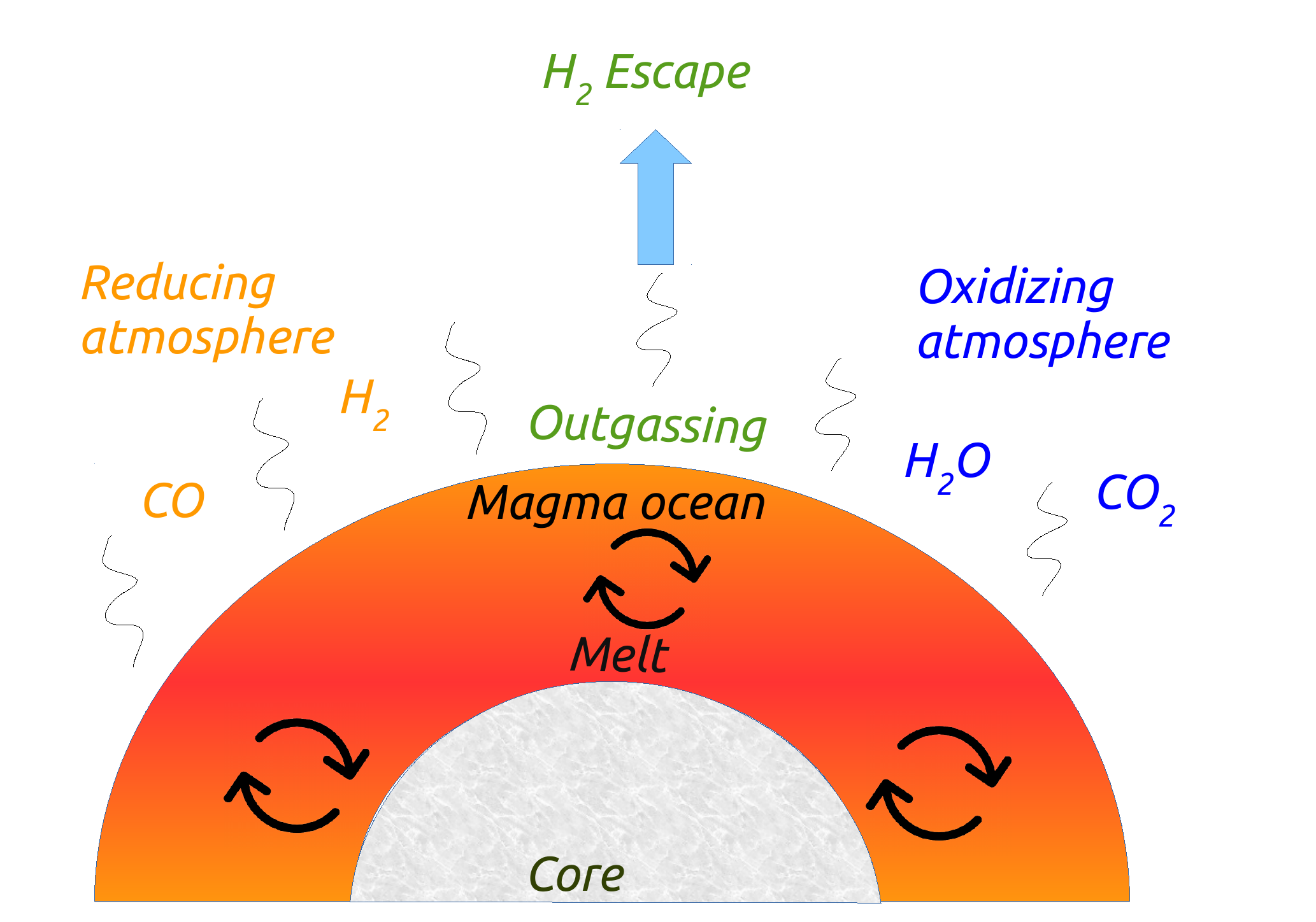}
 \centering
 \caption{Schematic showing coupled interior-atmosphere exchange via outgassing of reduced species such as \ch{H2} and CO (orange) for a reduced mantle (e.g., IW buffer) and oxidized species such as \ch{H2O} and \ch{CO2} (blue) for an oxidized mantle (e.g., QFM buffer) as the magma ocean solidifies. The \ch{H2} in the atmosphere is lost to space via escape. }
 \label{schematic}
\end{figure}

\begin{figure}[!hbt]
 \includegraphics[width=0.35\textwidth]{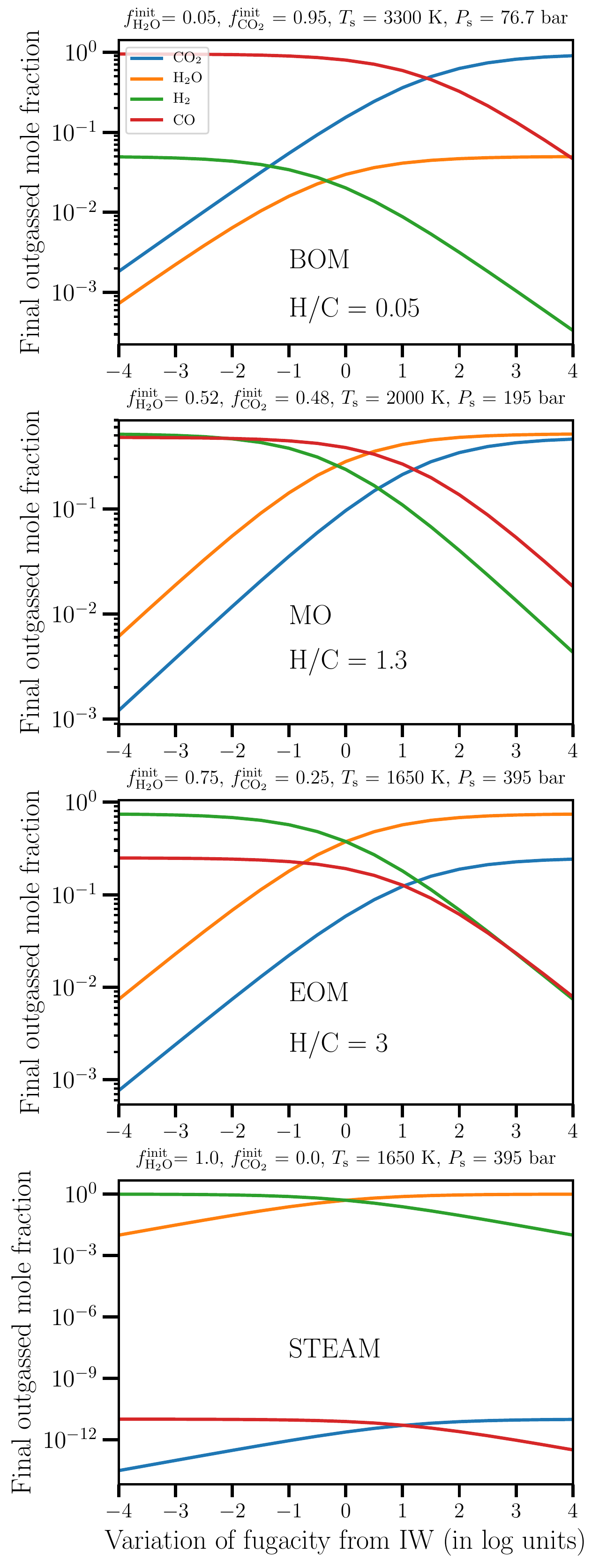}
 \centering
 \caption{Volatile chemical speciation in terms of outgassed mole fraction vs. oxygen fugacity of the mantle. The x-axis indicates the oxygen fugacity range in logarithmic units relative to the IW buffer. For reference (in log units), buffers QIF = IW-1, QFM = IW+3.8, and NiNiO = IW+4.2 approximately \citep{Wood1990}. From top to bottom, we show the BOM, active MO phase, EOM and a steam-atmosphere phase that are characterized by different initial H/C between hydrogen and carbon. The corresponding input values of the initial volume mixing ratio of volatiles, surface pressures $P_{\rm s}$ , and temperatures $T_{\rm s}$ \citep{Nasia2019} for the speciation model are indicated at the top of each panel.}

\label{volatile}
\end{figure}

A schematic for interior-atmosphere coupling is presented in Figure~\ref{schematic}. We started by taking the variation in surface pressure $p_{\rm s}$, surface temperature $T_{\rm s}$ , and volatile abundances $f^{\rm init}_{\ch{H2O}}$ and $f^{\rm init}_{\ch{CO2}}$  from the interior model output of \citet{Nasia2019}, that is, a mantle temperature of 4000 K and an initial mantle composition of  $X_{\ch{H2O}} = 0.05$ weight percent (wt\%) (550 ppm) and $X_{\ch{CO2}}= 0.01$ wt\% (130 ppm) as the input. We then applied the volatile speciation model from \citet{Ortenzi2020} as described in Sect.~\ref{s2} to obtain the final outgassed molar abundance of the species as a function of oxygen fugacity f\ch{O2} relative to the given mineral buffer (IW), as shown in Figure~\ref{volatile}.  This figure shows that for a reduced mantle (between IW and IW-4) and increasing input H/C from top to bottom, we obtain atmospheres that are rich in CO, CO+\ch{H2} mixtures, and \ch{H2}. On the other hand, for a more oxidized mantle with typically higher fugacity values (between IW and IW+4) and increasing input H/C from top to bottom (Fig.~\ref{volatile}), we obtain an atmosphere that is rich in \ch{CO2}, \ch{H2O}+\ch{CO2} , and \ch{H2O}. The bottom panel of Fig.~\ref{volatile} corresponds to an initially assumed 100\% \ch{H2O} which results in a almost pure \ch{H2} atmosphere for a reduced mantle and a pure \ch{H2O} atmosphere for an oxidized mantle. 

The initial outgassed volatiles \ch{H2O} and \ch{CO2} taken from  \cite{Nasia2019} with a mean molar mass $\mu_{\rm v}$ react with the melt and result in a different composition of the species outgassed to the atmosphere as calculated from the speciation model  (Section~\ref{s2}). When the number of moles of the species C-H-O is kept constant in the speciation model,  a different atmosphere with a new mean molar mass $\mu_{\rm atm}$ results, which has a different pressure at the bottom (surface) of the atmosphere, which we now call $p_{\rm boA}$. This pressure is calculated based on the volatile mass balance \citep{Bower2019},
\begin{equation}
    p_{\rm boA} = p_s \; \left (\frac{\mu_{\rm atm}}{\mu_{v}} \right),
    \label{pboA}
\end{equation}
where $p_s$ is the initial (surface) pressure of the volatiles calculated as a mono-gas atmosphere. 

Starting from the mass of the outgassed species \ch{H2O}  given by $M=P \mbox{*}A/g$, we obtained the following equation for the initial \ch{H2O} outgassing rate from the interior.
\begin{equation}
   R_{\ch{H2O}}^{\rm init} = \frac{1}{A}\frac{dN}{dt} =\frac{ N_A \; 1000}{g \; m_{\ch{H2O}}}\left(\frac{dP}{dt}\right)  \rm  \,\ molecules \,\ m^{-2} s^{-1},
    \label{molecules}
\end{equation}
where $dN$ is the number of \ch{H2O} molecules outgassed in a time interval $dt$, $N_A$ is Avogadro's number (6.022 $\times 10^{23}$), $A$ is the area of the Earth surface ($5.1\times 10^{14} \rm  \,\ m^{2}$), m$_{\ch{H2O}}$ is the molecular weight (18 g/mol), $g=9.81$ m/s$^{2}$ is the surface gravity, and $dp$ is the difference in outgassed pressure at a time interval $dt$ taken from coupled model result of \cite{Katyal2019} and \cite{Nasia2019}.

To obtain the outgassing rate of \ch{H2}, we used Eq.~\eqref{rH2} and Eq.~\eqref{molecules}. To account for the redox variation of the mantle and the volatile mass balance, we assumed that the initial rate of \ch{H2O} outgassing $R^{\rm init}_{\ch{H2O}}$  in Eq.~\eqref{molecules} is the total outgassing rate of combined (\ch{H2}+\ch{H2O}). Therefore, $R^{\rm init}_{\ch{H2O}} = r_{\ch{H2}} + r_{\ch{H2O}}$.  Inserting Eq.~\eqref{rH2} here, we obtain the rate of \ch{H2} outgassing as
\begin{equation}
    r_{\ch{H2}} =\frac{ R^{\rm init}_{\ch{H2O}}}{1 + R_1} \rm \,\ molecules \,\ m^{-2} s^{-1}.
    \label{H2final}
\end{equation}

\subsection{Convective lapse rate}\label{conv}

The convection  takes into account the heat transport by adiabatic expansion of a mixture of a nonideal condensable gas \ch{H2O} (because the critical point of water at $T_{\rm c} = 647$ K and $p_{\rm c} = 220$ bar is close to the T-p range considered in this study), denoted with subscript $c,$ and ideal, noncondensable gases (mixture of CO$_2$, CO, and H$_2$)  denoted with the subscript $d$. 
For a saturated water vapor atmosphere, the pressure exerted by the condensable is given by the Clausius-Clapeyron equation,
\begin{equation} \label{cc}
\frac{d p_s}{dT} = \frac{L_c(T)}{RT^2} p_s,
\end{equation}
where $p_s$ is the saturation vapor pressure and $L_c(T)$ is the latent heat of vaporization of the condensable, that is, \ch{H2O} in this case. Inserting this expression for $d\ln p_c/d \ln T$ into the formula given in \cite{Pierrehumbert2010} leads to the pseudoadiabatic slope as used in this study, which is given by \cite[see also][]{Ding2016}
\begin{equation}
\label{main2}
\frac{d \ln P}{d \ln T}= \frac{p_{\rm s}}{P}\frac{ L_c(T)}{R_c T}+\
                \frac{p_d}{P}\frac{c_{pd}}{R_d}\frac{1+\left(\frac{c_{pc}}{c_{pd}}+\left(\frac{L}{R_c T}-1 \right)\frac{L}{c_{pd} T} \right)r_{\rm sat}}{1+\frac{L}{R_d T}r_{\rm sat}},
\end{equation}
where $c_{pd}$ and $c_{pc}$ are the specific heat capacities of the noncondensable and condensable species, respectively. $r_{\rm sat}$ is the saturation mass mixing ratio, given as \citep{CatlingKastingBook2017}
\begin{equation}
 r_{\rm sat} = \epsilon f_{\rm sat}(\ch{H2O}) =  \epsilon \frac{p_{sat}}{P}.
\end{equation}
Here, $\epsilon$ is the ratio of mass of condensable and noncondensable species, $P$ is the total pressure, and $f_{\rm sat}$ is the volume mixing ratio of condensable at saturation. 
Eq.~\eqref{main2} reduces to Eq. (A13) of \cite{Kasting1988} at $r_{\rm sat} \to 0$. In this case, the atmosphere is dry and the temperature profile is obtained by  the dry adiabatic lapse rate given by (from Eq.~\ref{main2})
\begin{equation}
\label{pdry}
 \frac{d\ln p_d}{d \ln T}=\frac{c_{pd}(T)}{R_d},
\end{equation}
where $R_d$ is the gas constant for the noncondensable or the dry component of the mixture and is  obtained by Eq.~\eqref{rmix} of Appendix~\ref{App1}. When more than one noncondensable species is present, the mean molecular weight of the mixture of noncondensable is accounted for. 

At temperatures that are high enough and for a condensable species reservoir much larger than the atmospheric pressure exerted by the dry gases such that $r_{\rm sat} \gg 1$, $P \to p_{\rm sat}$ and $p_d/P \to 0$, Eq.~\eqref{main2} reduces to the Clausius-Clapeyron relation \citep{Pierrehumbert2010}, which is valid for an ideal gas equation. The temperature dependence of the latent heat of vaporization $L_c (T)$ and saturation vapor pressure $p_s$ of water cannot be neglected. For this, we used one of the most accurate formulations for calculating the saturation vapor pressure of water known as the Tetens formula \citep{Tetens1930}. This formula provides a very good approximation for the saturation vapor pressure of water with an error lower than $\sim1$\% \citep{Huang2018} in the temperature range 0-100~$^\circ$C and is given as follows:

\begin{align}
\label{tetens}
%
     p_s (T)= 
\begin{cases}
    p_s^{\rm ref} \exp \left( \frac{17.625T}{T+238.3}  \right) ,& \text{if } T\geq 0\\
   p_s^{\rm ref} \exp \left( \frac{21.875T}{T+265.5}  \right)   ,              & \text{otherwise}.
\end{cases}
\end{align}
Here $p_s^{ref} = 610.78 \, \rm Pa$ is the reference water vapor pressure at the triple point, and $T$ is given in $^{\circ}$C and equal to $T$- 273 for $T$ in units of Kelvin.

\begin{figure}[!hbt]
  \includegraphics[width=\hsize]{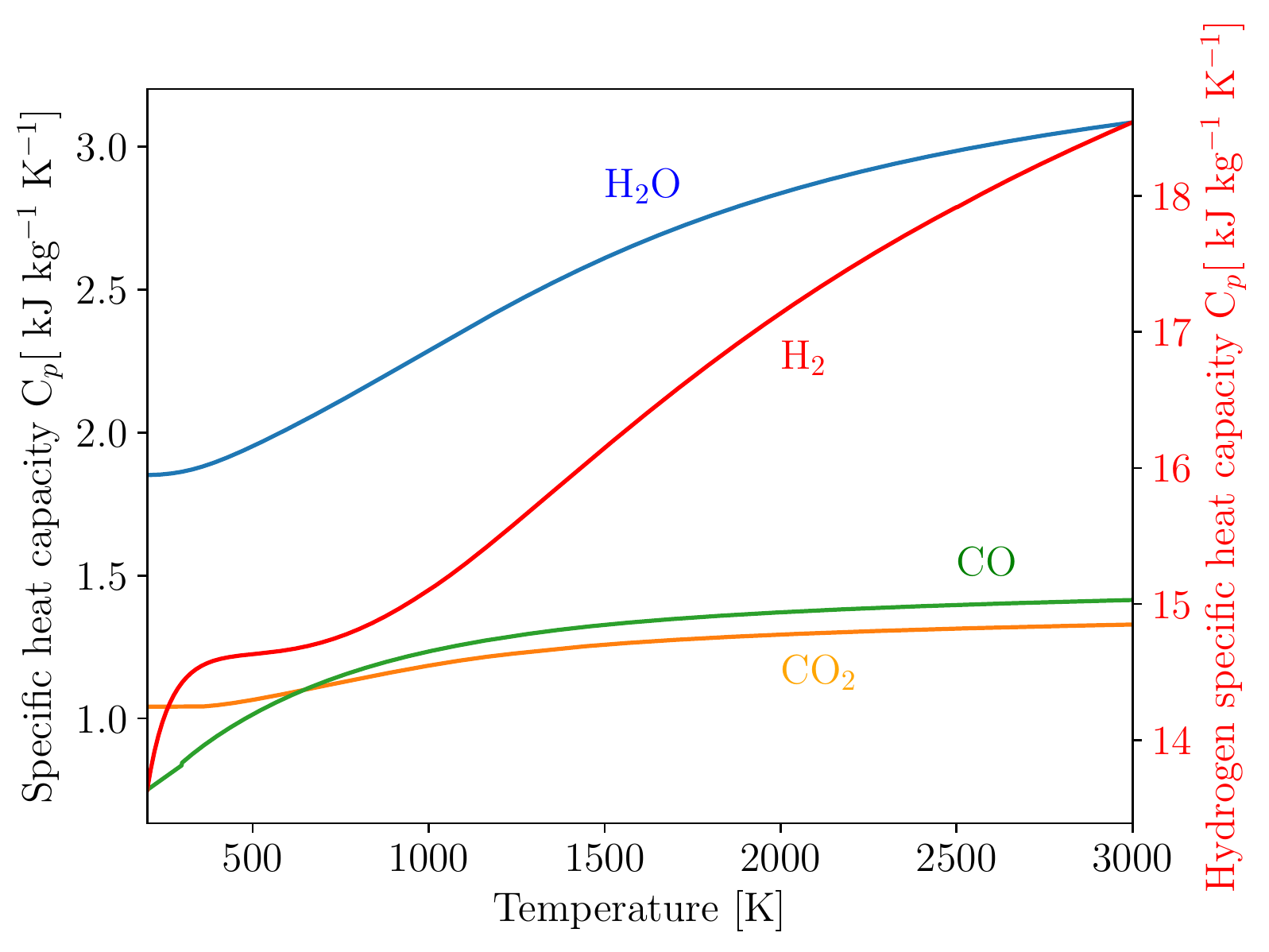}
 \centering
 \caption{Specific heat capacities as a function of temperature for volatile species \ch{H2O}, CO, and \ch{CO2} on the left x-axis and \ch{H2} on the right x-axis as obtained from the Shomate equation described in the main text.}
 \label{CP}
\end{figure}

Next the term $c_{pc}$ in Eq.~\eqref{main2} is the temperature-dependent specific heat capacity of water adapted from \cite{Wagner2002} and applicable to high-temperatures conditions given by \citep[also see][]{Katyal2019}
\begin{equation}
\frac{c_{pc}(T)}{R_c} = 1+n_{3}+ \sum_{i=4}^{8}n_i \frac{(\gamma_i \tau)^{2}\exp(-\gamma_i \tau)}{[1-\exp(-\gamma_i \tau)]^{2}}, \label{e7}
\end{equation}
where $\tau=T_{\text c}/T$ and $T_{\text c}$ is the critical temperature (647 K). 
The values of the coefficients $n_{i}$ and $\gamma_{i}$ are taken from Table 6.1 of \cite{Wagner2002}. The total heat capacity of the dry mixture $c_{pd}$ is calculated as a linear combination of individual heat capacities of the noncondensable species weighted by their volume mixing ratios. The total $c_p$ is thus the sum of specific heat capacity of moist and dry component, weighted by the respective volume mixing ratios, and is given by
\begin{equation}
	c_p(T)= \frac{\sum_i x_{di} c_{pdi}(T)+ \sum_i x_{vi} c_{pci}(T)}{\sum_i x_{di} + \sum_i x_{ci}},
\end{equation}
which is valid for temperatures in the range 200-2000 K. The denominator in this equation is equal to 1 for our case. The specific heat capacity of dry gases such as CO$_2$, CO, and H$_2$ is derived from the Shomate equation \citep{shomate1940} in units of $\rm kJ/kg\; K$,
\begin{equation}
 c_p = A + BT + CT^2+DT^3+E/T^2,
\end{equation}
where the coefficients $A, B, C, D,\text{and } E$ are taken from \cite{janaf1} and $T= \rm Temperature/1000 $ K for the temperature range valid between 298-6000 K. For temperatures below 298 K, the specific heat capacities of CO, \ch{CO2} , and \ch{H2} are taken from analytical expressions based on a least-squares fit of data from \cite{Lide2000}. The specific heat capacity as a function of temperature in the range 200-3000 K for the four gases we considered is shown in Fig.~\ref{CP}. 

In order to obtain the temperature profiles, we followed the same procedure as in \cite{Katyal2019} , that is, for surface temperatures $T_s > 647$ K, we used the dry adiabatic lapse rate (Eq.~\ref{pdry}) by assuming a fixed surface temperature and pressure at the bottom of the atmosphere (Eq.~\ref{pboA}). The temperature profile follows the moist adiabatic lapse rate (second term in  Eq.~\ref{main2}) when the dry adiabatic lapse rate intersects with the moist adiabatic lapse rate. 
The
related work of \citet{Katyal2019} calculated temperature profiles with only one gas species, that is, H$_2$O, in the atmosphere. We have 
additionally included CO$_2$, CO, and H$_2$ calculated from the volatile speciation model described in Sect.~\ref{s2}. Our model accounts for the convective processes in the atmosphere up to an altitude corresponding to the top of the atmosphere pressure $p_{\rm ToA}=0.1$ Pa  (similar to \cite{Pluriel2019}). The atmospheric layers are vertically spaced in log pressure coordinates. When the temperatures in the dry/moist adiabat fall below 200 K, the temperature structure in our calculations follows an isothermal profile with a constant tropospheric temperature of 200 K.

\subsection{Radiative transfer code}\label{rad}
The  generic atmospheric radiation line-by-line infrared code \cite[GARLIC;][]{Schreier2014} was used to compute thermal emission and transmission spectra from 10 to 30,000 cm$^{-1}$ with p,T and the composition of the atmosphere as the input. A detailed description of the radiative transfer code is also available in \cite{Katyal2019}. For the verification and validation of GARLIC, see \cite{Schreier2018b,Schreier2018a}. The absorption coefficients were calculated  line-by-line (lbl)  for \ch{H2O}, \ch{CO2}, \ch{H2}, and CO from the HITRAN2016 database \citep{Gordon2017}. Table~\ref{continuum} shows the list of various molecules and sources for the continuum available and relevant for this study.

\begin{table}
\centering
    \caption{List of continua from the HITRAN CIA list\tablefootmark{a} \;and other sources.
    \label{continuum}}
\begin{tabular}{c c}
\hline\hline
    {Molecule} & {Continuum}\\
\hline
         H$_2$ -- H$_2$ & \citet{Abel2011} \\
         CO$_2$ -- CO$_2$ & \citet{Baranov2018} \\
         H$_2$O $\rm_{self}$ & MT\_CKD\_2.5\tablefootmark{b} \\
         H$_2$O $\rm_{foreign}$ & MT\_CKD\_2.5\tablefootmark{b} \\
         CO$_2$ $\rm_{foreign}$ & 
         MT\_CKD\_2.5\tablefootmark{b}\\
\hline 
\end{tabular}
    \tablefoottext{a}{\url{www.hitran.org/cia/} \citep{karman2019}}
    \tablefoottext{b}{\url{http://rtweb.aer.com/continuum_frame.html} \citep{mlawer2012}}
\end{table}

\begin{table}
\centering
    \caption{List of measured reference Rayleigh cross sections $\sigma_{0,i}$ and central wavelengths $\lambda_{0,i}$ for the molecules.
    \label{RayLeigh}}
\begin{tabular}{c c c c}
\hline\hline
    {Molecule} & {$\sigma_{0,i}$ [cm$^2$]} & $\lambda_{0,i}$ [$\mu$m] &
    {Reference}\\
\hline        
    \ch{CO2}&$12.4 \cdot 10^{-27}$&0.53224& \cite{Sneep2005}\\
 \ch{CO}&$6.19 \cdot 10^{-27}$&0.53224&\cite{Sneep2005}\\
 \ch{H2}&$1.17 \cdot 10^{-27}$&0.5145&\cite{Shardanand1977}\\
\hline
\end{tabular}
\end{table}

 Rayleigh scattering by molecule \ch{H2O} was obtained using the formalism described in \cite{Murphy1977} \citep[also see][]{Markus2020}. For other molecules such as \ch{CO2}, CO, and \ch{H2},  we have used the Rayleigh cross sections using the formalisms of \cite{Sneep2005} and \cite{Shardanand1977} across the whole spectrum in units of $\rm cm^{2}$/molecules expressed as
\begin{equation}\label{sigma}
\sigma_{\rm {Rayleigh}, i}(\lambda) = \sigma_{0,i}\;\left(\frac{\lambda_{0,i}}{\lambda}\right)^{\alpha}.
\end{equation}
 Here, $\alpha=4$, $\sigma_{0,i}$ and $\lambda_{0,i}$ are the reference cross sections and wavelengths for the various molecules used in this study that are listed in Table~\ref{RayLeigh}.

 To mimic optically thick clouds in our simulations (Section~\ref{clouds1}), we assumed clouds were 
present throughout the atmosphere \citep{Kalt2009} with the uppermost cloud deck coinciding with the lowermost $T=200$ K altitude. The cloud-scattering cross section is expressed in a similar formalism as Eq.\eqref{sigma}, but with $\alpha=0$. This therefore acts as a scaling term with no wavelength dependence. For reference, the terrestrial aerosol-scattering cross
section is about $10^{-27} \rm \, cm^2$ \citep{Sarah2018}. That study \citep{Sarah2018}  ruled out haze-scattering cross sections smaller than 9 $\times 10^{-25}$ cm$^{2}$ to 3$\sigma$  and 3 $\times$ 10$^{-23}$ cm$^{2}$ to 1$\sigma$. Therefore we assumed a cloud cross section $\sigma_{\rm cloud} \equiv  \sigma_{0_i} = 6 \times \rm 10^{-25} \rm \, cm^{2}$ , that is, $6 \times 10^{2}$ times larger than the Earth-scattering cross section in our simulations as presented in Section~\ref{clouds1}.

    

The aerosol optical depth integrated along the path is then given by
\begin{equation}\label{cloud}
    \tau_{\rm aerosol} = \sigma_{\rm aerosol} \cdot N_z, 
\end{equation}
where $N_z$ is the column density of the molecular species in units of $\rm molecules \, cm^{-2}$ \cite[see also][]{Kalt2009,Yan2015}. 



 As a first step, we performed convective-radiative
calculations without coupled chemistry similar to previous modeling studies of the magma ocean period. For future work, we plan to include the effect of chemical equilibrium in the lower atmospheric layers by applying the self-consistent convection-climate-photochemistry column model \citep{Markus2020,Fabian2020} to the magma ocean period in order to study the effect upon transmission and thermal emission spectra and atmospheric escape, for example.

\subsection{Transmission spectra}

The volatile chemical speciation model along with the atmospheric model described in Sections~\ref{s2}-\ref{rad} applicable to the early Earth magma ocean phase can be used to characterize exoplanets in the magma ocean phase \citep[see Figure 12 of][]{Nasia2019}, for instance, to detect the spectral features and obtain the vertical extent of the atmosphere observationally. We used a well-established diagnostic tool, transmission spectroscopy, to probe the molecular composition of exoplanetary atmospheres. It provides the effective height of the atmosphere as a function of wavelength \cite[see, e.g.,][]{benekeseager}. 

In GARLIC, the theoretical transmission spectrum is expressed as the effective height of the atmosphere given as follows \citep{Kalt2009,Schreier2018a,Katyal2019}:
\begin{equation}
    h(\lambda) = \int_0^\infty A(z,\lambda) \; dz = \sum_i A_i (\lambda) \; \Delta h_i,
    \label{height}
\end{equation}
where $A_i=1-\mathcal{T}_i$ is the absorption along the $i$th incident ray with a transmission $\mathcal{T}_i$ that traverses through the (exo)planetary atmosphere tangentially at a height $h_i$ and continues to travel to the observer at Earth.

The atmospheric transit depth, $t_{\rm atm}$ is given by \citep{Fabian2020}
\begin{equation}\label{tdepth}
    t_{\rm atm}(\lambda)=\frac{(R_p+h(\lambda))^2}{R_s^2}-\frac{R_p^2}{R_s^2},
\end{equation}
where $R_{\rm p}$ is the planetary radius, and $R_{\rm s}$ is the stellar radius.

\subsection{Atmospheric escape}\label{escape_basic}
For the atmospheric escape of \ch{H2}, we considered hydrodynamic fluxes of escaping \ch{H2}, possibly diffusion-limited at the homopause $\sim100 \; \rm km$, in the presence of a static background atmosphere, similar to the approach of \cite{Kuramoto2013,Catling2017} and \cite{Zahnle2019}. 
At higher levels above the homopause, where the escaping \ch{H2} gas is accelerated as a result of the absorption of stellar high-energy photons, hydrogen exists mainly in atomic form.
For the quantitative implementation of the escape fluxes, we relied on \cite{Zahnle2019}, who solved the hydrodynamic, possibly diffusion-limited, problem in the atmosphere. In particular, we used their Eq. 3, which parameterizes the transition from energy-limited escape to diffusion-limited escape. Diffusion-limited escape becomes effective when the atmosphere is more strongly irradiated and the escaping hydrogen cannot be replenished from below the homopause.


The diffusion-limited flux $\phi_{\rm dl}$ generally provides an upper limit to the escape of lighter species such as hydrogen in a background heavier and stationary gas. The escape rate for \ch{H2} can be written as \citep{Hunten1973}
\begin{equation}
\label{DLflux}
 \phi_{dl}= b_{aj} \; f_{\rm tot}(\ch{H2})\left(\frac{1}{H_a}-\frac{1}{H_{\ch{H2}}}\right),
\end{equation}
where $b_{\rm aj}$ is the binary diffusion coefficient between the background heavier gas $a$ such as \ch{CO2}, \ch{CO} and \ch{H2O} and the escaping gas \ch{H2}. Here, $f_{\rm tot}\rm(H_2)$ is the hydrogen volume mixing ratio (VMR) at the homopause. $H_a$ and $H_{\ch{H2}}$ represent the unperturbed scale heights ($=kT/mg$) of the heavier gas $a$ and the escaping gas \ch{H2} at the homopause, respectively. We consider \ch{H2} as the dominant H-bearing species, that is, $f_{\rm tot} \equiv f(\ch{H2})$. The binary diffusion coefficients $b_{aj}$ for H$_2$ in CO and \ch{CO2} are roughly similar and taken to be $3 \times 10^{21} \rm \,\ m^{-2} \,\ s^{-1}$ \citep{Marrero1972}. 
We validated the escape rate obtained for the current Earth using Eq.~\eqref{DLflux}. The total hydrogen VMR was taken to be $f_{\rm tot}\rm(H_2) = 7.15$ ppm with the background-dominant gas as \ch{N2}. The diffusion coefficient between the \ch{H2} and \ch{N2}, $b_{aj}$ of $1.7 \times 10^{21} \rm \,\ m^{-2} \,\ s^{-1}$ was taken from \cite{Hunten1973}. We verified that the diffusion-limited escape of \ch{H2} is $\sim 1.5 \times 10^{12} \,\ \rm \ch{H2} \,\  \rm molecules \,\ \rm m^{-2} \,\ \rm s^{-1}$ , as also reported by \cite{Hunten1976}. In terms of the loss of H-atoms to space, this is equivalent to a mass-loss rate of $\sim$3 kg/s.

The energy-limited flux was obtained as the ratio of stellar XUV energy incident upon the planet and the energy required to lift a given mass out of the Earth potential well and into the space. The mass-loss rate of hydrogen using the energy-limited formula \citep{Watson1981} for hydrodynamic escape \citep[also see][]{koskinen2014,Hamano2015} is given by
\begin{equation}
    \dot M_{\rm el} = \frac{\pi \epsilon R_p R_{\rm XUV}^{2} F_{\rm XUV}(t)}{GM_p}    \; \rm  g \,\ s^{-1},
    \label{el}
\end{equation}
where $R_{\rm XUV}$ is the
radial distance at which the XUV energy from the star is deposited, and satisfies
$R_{\rm XUV}$ > $R_p$, and $\epsilon$ is the heating efficiency < 1. Upon assuming $R_{\rm XUV} \equiv R_p$ \citep{Zhang2020}, this equation can be written as
\begin{equation}
   \dot M_{\rm el} = \frac{\pi \epsilon R_p^3 F_{\rm XUV}(t)}{G M_p}  \; \rm  g \,\ s^{-1}.
    \label{el1}
\end{equation}
Here, $R_{\rm p}$ is the Earth radius (6.4 $\times$ 10$^{8}$ cm), $M$ is the mass of Earth (5.4 $\times$ 10$^{27}$ gm), $G$ is the gravitational constant (6.67 $\times$ 10$^{-8}$ cm$^{3}$ \rm g$^{-1} \rm s^{-2}$ ), 
$F_{\mathrm{XUV}}(t)$ is the time-evolving flux from the host star at XUV wavelengths at 1 AU obtained by the relation $F_{\mathrm{XUV}}(t)= 5(4.5/t)^{1.24}$ in units of $\rm \; erg \; cm^{-2} s^{-1}$, where $t$ is the age of the Sun in billion years \citep{Zahnle2019}, and $\epsilon$ is taken to be 0.5 to be consistent with \cite{Zahnle2019}. Owing to the fast temporal evolution in the XUV output of the young Sun (related to whether it was a fast, moderate, or slow rotator), the $S$ value defined as $S(t) \equiv F_{\mathrm{XUV}}/F_{\mathrm{XUV}\odot}$ can range between 10 and 100  during the Hadean \citep{Tu2015,Lammer2018}. The lifetime of \ch{H2} atmospheres on early Earth is calculated as the ratio of mass-loss rate (Eq.~\ref{el1}) to the total mass of \ch{H2} in grams, $m_{\ch{H2}}$  , in the atmosphere as
\begin{equation}
    t \sim \left[\frac{G}{\epsilon \pi}\right] \left(\frac{M_p}{R_p^{3}}\right) \frac{m_{\ch{H2}}}{F_{\rm XUV}}.
    \label{timescale}
\end{equation}

 \cite{Zahnle2019} have shown that the diffusion-limited escape applies to conditions of higher levels of irradiation, whereas energy-limited escape may be a better approximation for lower levels of irradiation. Building upon their findings, we assumed that the flux is given by the expression 
 \begin{equation}
    \phi_{\ch{H2}} = \frac{1 \times 10^{12} f_{\ch{H2}}S}{\sqrt{1+0.006 S^2}}   \,\  \rm molecules \,\  cm^{-2}  s^{-1},
	 \label{zahnlebestfit}
 \end{equation}
 which is a good fit to their figure 5 if the VMR of \ch{H2} is lower than 0.2. 

The hydrogen mass-loss $dM$ (in grams) due to escape at a particular time step $dt$ is approximated as follows:
\begin{equation}
    dM = 4\pi R_p^2\;   \phi_{\ch{H2}} m_{\ch{H2}} \; dt.
\label{dm}
\end{equation}
Here, $m_{\ch{H2}}$ is the mass of \ch{H2} in grams per molecule, and $\phi_{\ch{H2}}$ is the escaping flux from Eq.~\eqref{DLflux} or Eq.~\eqref{zahnlebestfit}. 

 The actual radius contribution in Eq.~\eqref{el1} and Eq.~\eqref{dm} comes from the sum of the radius of the planet and height of the exobase, that is, ($R_p$+$H_e$). The height of the exobase, $H_e$ , is given by the altitude level where the Lyman-edge opacity of hydrogen becomes 1. For the scenario presented in Section~\ref{escape}, we obtain that ($R_p$+$H_e) \sim R_p$ (calculation not shown here).

\section{Volatile outgassing scenarios}\label{sce}
 We used three main scenarios in this work. Scenario 1 is for a  surface temperature of $3300 \rm K$ and surface pressure of 76.7 bar, that is, at the beginning of the magma ocean (BOM) period, as shown in Figure~\ref{volatile} (top) and Table~\ref{buffer1}. Scenario 2 is similar to scenario 1, but for a surface temperature of 1650 K and pressure of 395 bar, that is, at the end of the magma ocean (EOM) period, as shown in Figure~\ref{volatile} and Table~\ref{buffer2}. The conditions where the MO phase ends are $T_{\rm s}=1650$ K at a solidification timescale of $\sim$ 1 Myr  according to the coupled interior-atmospheric model results of \cite{Katyal2019} and \cite{Nasia2019}. Previously, a coupled atmospheric-interior model by \cite{Salvador2017}, which was applied around the MO phase, delivered P-T conditions with a similar \ch{H2O}-\ch{CO2} composition as this study.
 
 Scenario 3 is for representative cases taken 
 from the study of \cite{Pluriel2019}, as shown in Table~\ref{buffer} referring to an \ch{H2O} dominated and a \ch{CO2} dominated atmosphere. Scenarios 1 and 2 cover a representative range of P, T, and mantle oxidation state during the MO period. The buffers in the three scenarios were chosen to represent the mantle redox state, ranging from reducing to oxidizing. 

Our choice of three scenarios (with varying initial volatile abundances of \ch{H2O} and \ch{CO2}) and varying mantle fugacity, hence leading to different atmospheric compositions, is motivated by previous works in the literature that also explored the effect of composition by varying key species such as \ch{H2}, CO, \ch{CO2}, \ch{H2O}  for magma ocean studies \citep{Katyal2019,Nasia2019,Bower2019}, Venus-type planets with \ch{CO2} dominated atmosphere \citep{Hamano2013,Pluriel2019}, pure steam-dominated atmosphere \citep{Katyal2019,Schaefer2016,Hamano2015}, and the effect of adding \ch{H2} to the atmosphere of early Mars \citep{Ramirez2014}.

 
%

\begin{table*}[!hbt]
    \centering
    \caption{Scenario 1 (BOM) for the three cases (1.1, 1.2, and 1.3). Columns 2 and 3 show the initial assumed mole fractions arising from \ch{H2O} and \ch{CO2} outgassing, which is the input to the speciation model. The assumed buffer values for strongly reducing (IW-4),  reducing (IW),
 and highly oxidizing (IW+4) have been used to study the effect of speciation under these conditions. The four columns on the right shows final outgassed species from the speciation model. For these scenarios, the surface p, T setting is fixed to be $T_s = 3300$ K and $P_{\rm s}=76.7$ bar. The $p_{\rm boA}$ calculated from the new molecular weight of the atmosphere is also shown. Scenario 1.1 with IW+4 case resembles the  p,T during the BOM as obtained by \cite{Nasia2019}. The most dominant species in the atmosphere are marked in bold for each of the cases. }
\begin{tabular}{|c||c|c|c|c||c|c|c|c|c||}
 \hline
 Scenario&\multicolumn{2}{c|}{Initial outgassing} &$P_{\rm s}$ (bar)&Buffer&$p_{\rm boA}$ (bar)&\multicolumn{4}{c|}{Final  outgassing} \\
 & $f^{\rm init}_{\rm \ch{H_2O}}$&$f^{\rm init}_{\rm \ch{CO2}}$&&&&$f_{\rm \ch{CO2}}$& $f_{\rm \ch{H2O}}$&$f_{\rm \ch{H2}}$&$f_{\rm CO}$\\
 \hline
 \hline
1.1&0.05&0.95 &76.7& IW-4&48&0.0018 &0.0007& 0.049& {\bf 0.94}\\
&&&&IW&53.2&0.15& 0.029& 0.02&{\bf 0.79}\\
& &&&IW+4&75.3&{\bf 0.90}& 0.049& 0.000& 0.046\\
\hline
1.2&0.75&0.25 &76.7&IW-4& 27.1&0.000& 0.01& {\bf 0.74}& 0.25\\
&&&&IW&51&0.040& {\bf 0.44}& 0.30& 0.20\\     
& &&&IW+4&75.8&0.24& {\bf 0.74}& 0.005& 0.0122\\
\hline
1.3&1.0&0.00  &76.7&IW-4&9.5&0.0 &0.014 &{\bf 0.98}&0.00\\
&&&&IW&49.1&0.00&{\bf 0.60}& 0.40 &0.00\\
& &&&IW+4&76.2&0.00&{\bf  0.99}& 0.0069& 0.00\\
 
 \hline
\end{tabular}


\label{buffer1}
\end{table*}

\begin{table*}[!hbt]
    \centering
    \caption{Same as Table~\ref{buffer1}, but for scenario 2 (EOM), investigating the effect of speciation under these conditions. The surface p,T setting is fixed at  $T_s = 1650$ K and $P_s = 395$ bar. Scenario 2.2 with IW+4 resembles the  p,T at the EOM, as obtained by \cite{Nasia2019}.}
    \begin{tabular}{|c||c|c|c|c||c|c|c|c|c||}
 \hline
 Scenario&\multicolumn{2}{c|}{Initial outgassing } &$P_{\rm s}$ (bar)&Buffer&$p_{\rm boA}$ (bar)&\multicolumn{4}{c|}{Final  outgassing} \\
 & $f^{\rm init}_{\rm \ch{H_2O}}$&$f^{\rm init}_{\rm \ch{CO2}}$&&&&$f_{\rm \ch{CO2}}$& $f_{\rm \ch{H2O}}$&$f_{\rm \ch{H2}}$&$f_{\rm CO}$\\
 \hline
 \hline
2.1&0.05&0.95 &395&IW-4&247.5& 0.003 &0.0005& 0.049& {\bf 0.94}\\
&&&&IW&283.7&0.22& 0.025& 0.025&{\bf 0.72}\\
& &&&IW+4&390.4&{\bf 0.92}& 0.05& 0.000& 0.03\\
\hline
2.2&0.75&0.25 &395&IW-4& 139&0.0007& 0.007& {\bf 0.74}& 0.25\\
&&&&IW&248&0.058& {\bf 0.37}& {\bf 0.37}& 0.19\\     
& &&&IW+4&391&0.24& {\bf 0.74}& 0.007& 0.007\\
\hline
2.3&1.0&0.0 &395&IW-4&47.3&0.00 &0.009 &{\bf 0.99}&0.00\\
&&&&IW&219&0.00&{\bf 0.50}& {\bf 0.50} &0.00\\
& &&&IW+4&391.5&0.00&{\bf  0.99}& 0.0099& 0.00\\
 
 \hline
\end{tabular}

\label{buffer2}
\end{table*}

 \begin{table*}[!hbt]
    \centering
    \caption{Scenarios showing the output from the volatile speciation model applied to scenario 3.1, \ch{H2O}-dominated atmosphere and scenario 3.2, \ch{CO2}-dominated atmosphere at fixed $T_s=1500$ K for two different surface pressures.  The atmospheric pressure $p_{boA}$ shown above is recalculated based on the new molecular weight of the outgassed species. The initial and final outgassed species are shown as mole fractions.
The most dominant species in the atmosphere are marked in bold in each case.}
    \begin{tabular}{|c||c|c|c|c||c|c|c|c|c| c ||}
 \hline
 Scenario&\multicolumn{2}{c|}{Initial outgassing} &$P_{\rm s}$ (bar)&Buffer&$p_{\rm boA}$ (bar)&\multicolumn{4}{c|}{Final  outgassing}& Reference \\
 & $f^{\rm init}_{\rm \ch{H_2O}}$&$f^{\rm init}_{\rm \ch{CO2}}$&&&&$f_{\rm \ch{CO2}}$& $f_{\rm \ch{H2O}}$&$f_{\rm \ch{H2}}$&$f_{\rm CO}$&\\
 \hline
    \hline
3.1&0.66&0.33 &300&None&300& 0.33 &0.66& -& - & \cite{Pluriel2019}\\
&&&&IW&  191.8&0.083& 0.311& {\bf 0.35}& 0.24&\\
& &&&IW+4 & 297&0.32& {\bf 0.66}& 0.006& 0.007&\\
\hline
3.2&0.02&0.98 &510&None&510& 0.98& 0.02&- & -&\cite{Pluriel2019}\\
&&&&IW&371.2&0.25& 0.009& 0.01& {\bf 0.73}&\\     
& &&&IW+4 &504.7&{\bf 0.95}& 0.019 &0.0& 0.023&\\
\hline

\end{tabular}

\label{buffer}
\end{table*}

\section{Results}\label{s3}
\subsection{Temperature profiles}\label{tp}

\begin{figure*}[!hbt]
 \includegraphics[width=0.9\textwidth]{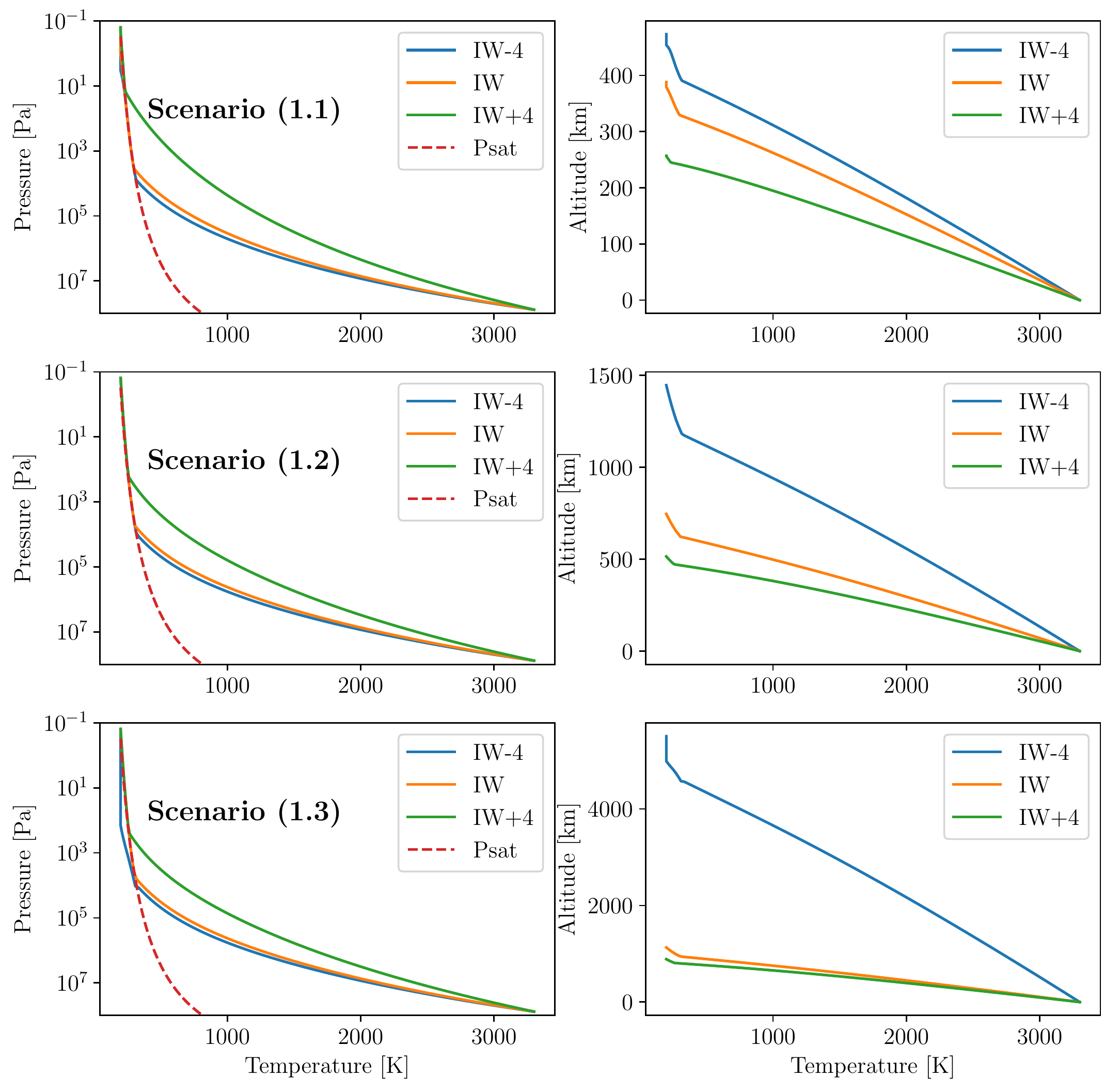}
 \centering
 \caption{{\it Left:} Pressure-temperature profiles for the BOM scenarios 1.1 (upper), 1.2 (middle), and 1.3 (lower panel) for three different buffers IW-4, IW, and IW+4. The dashed red line represents the saturation water vapor curve. Pressures at the bottom of atmosphere values are taken from Table~\ref{buffer1}. {\it Right}: Altitude-temperature  profiles for scenarios 1.1 (upper), 1.2 (middle), and 1.3 (lower panel).  We note the different altitude ranges for the panels on the right.}


 \label{redox_BOM}
\end{figure*}

\begin{figure*}[h]
 \includegraphics[width=0.9\textwidth]{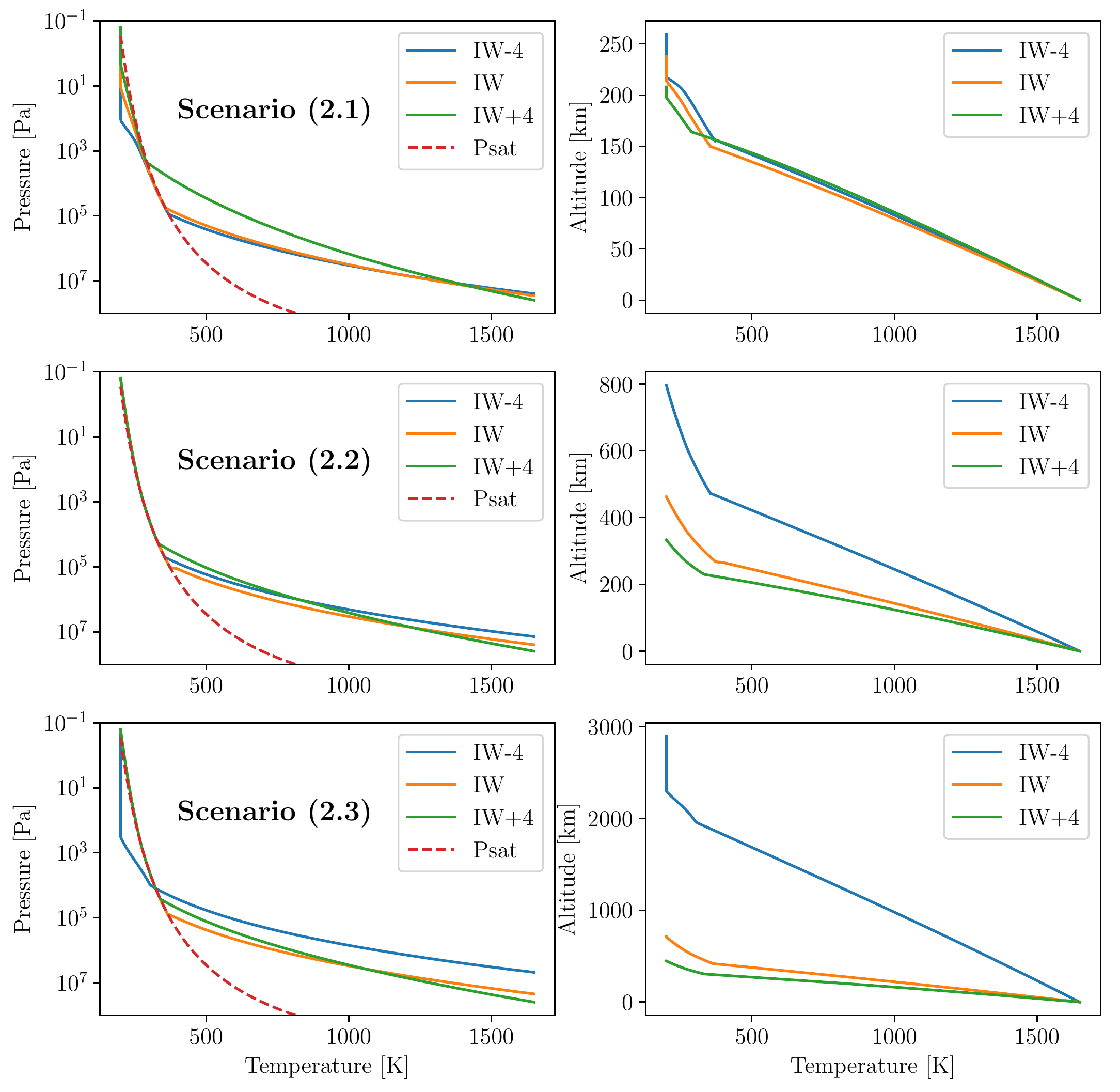}
 \centering
 \caption{  Same as Fig.~\ref{redox_BOM}, but for p, T conditions and scenarios as summarized in Table~\ref{buffer2}, representing the EOM.} 
  \label{redox_EOM}
\end{figure*}

Figures~\ref{redox_BOM} and ~\ref{redox_EOM} show the pressure-temperature (p-T) and altitude-temperature (z-T) profiles obtained using the convection lapse-rate formulation as described in Section~\ref{conv} for scenarios 1 and 2 as shown in Tables~\ref{buffer1} and \ref{buffer2}, respectively. In Figure~\ref{redox_BOM} scenario 1.1 (95\%\  initial CO$_2$, 5\%\  initial H$_2$O, top left panel, p-T), the results suggest a dry adiabat in the lower unsaturated troposphere that is
related to the high surface temperature. The green curve (oxidized buffer case) adiabat is steeper in this region than the blue and orange curves because it is CO$_2$-dominated, leading to a low heat capacity (Figure~\ref{CP}) and accordingly to  a steeper calculated lapse rate (equation~\ref{pdry}). This steep slope results in the green curve that intersects the saturated vapor curve (dashed red line) at lower pressures than for the blue and orange curves, which have lower CO$_2$ but more CO and H$_2$O (Table~\ref{buffer2}) owing to higher heat capacities (Figure~\ref{CP}) and therefore shallower dry lapse rates (equation~\ref{pdry}).

In Figure~\ref{redox_BOM} for z-T (top right panel), the green curve features the lowest atmospheric height (geometric thickness) due to its larger atmospheric molecular weight and therefore smaller scale height than the other, lighter atmospheres shown by the orange and blue curves. 
Above the dry adiabat regime (e.g., above about 320 km for the orange line), the slope steepens corresponding to the wet adiabat as temperatures are low enough to allow for condensation. In the uppermost atmosphere, a fixed iso-profile temperature ($T=200$ K) (e.g., occurring above about 380 km for the orange line) is imposed in our model for the radiative regime, thereby following other studies \citep{Lupu2014,Pluriel2019,Katyal2019}.    


In the middle and lower panels of Figure~\ref{redox_BOM}, the initial H$_2$O (CO$_2$) amounts are increased (decreased) (see scenarios 1.2 and 1.3 in Table~\ref{buffer1}), which leads to light, hydrogen-dominated atmospheres for the reduced buffer cases. Because H$_2$ has a heat capacity between that of H$_2$O and CO$_2$ {\bf or} CO (Figure~\ref{CP}), the green slope in the middle and lower left panels is now closer to the blue and orange slopes compared with the upper panel because the difference in heat capacities is now smaller.
In the middle and lower right panels, results suggest extended scale heights for highly reduced atmospheres with dominant \ch{H2} (e.g., scenario 1.3, IW-4) because its molecular weight is lower than that of oxidized species such as \ch{H2O} or \ch{CO2} in the atmospheres (e.g., scenario 1.1 and 1.2, IW+4) and subsequently the change in altitude range of the figure is notable.

Figure~\ref{redox_EOM} illustrates the EOM scenarios (Table~\ref{buffer2}). Compared with the earlier BOM case (Figure~\ref{redox_BOM}) with a surface temperature of 3300 K, the surface temperature for the EOM scenario has cooled to 1650 K and the surface pressure has increased from 76.7 bar (BOM) to 395 bar (EOM). The cooler temperatures generally favor the formation of chemical species with weaker bond dissociation energies, that is, H$_2$ (4.52 eV) over H$_2$O (5.1 eV) and CO$_2$ (5.51 eV) over CO (11.1 eV) (where eV denotes electron Volt), which is consistent with the changes produced by the chemical speciation model (compare Table~\ref{buffer2} with Table~\ref{buffer1}). Larger changes of up to a few dozen percent in the species mole fraction occur for scenario 2.2 compared with scenario 1.2, in which the relative amount of C-H-O (0.75:0.25 for H$_2$O:CO$_2$) is the most similar. Figure~\ref{redox_EOM} (EOM) shows modest changes in comparison to Figure~\ref{redox_BOM} (BOM), which are associated with the change in p,T  and corresponding changes in speciation and heat capacity, hence in the slope of the adiabats. The altitude decrease is mainly
associated with the decrease in temperature.

 Figures~\ref{redox_BOM} and \ref{redox_EOM} show that (1) the 
transition between dry and moist adiabat (slope change) always appears
to be located on the saturation curve $P_{\rm sat}$ and it effectively occurs at lower temperature than the saturation temperature (for similar pressure); and (2) for $f_{\ch{H2O}}$ $\geq$ 0.1 in the atmosphere, the moist adiabat always coincides with the water vapor saturation curve $P_{\rm sat}$.

\begin{figure}[t]
  \includegraphics[width=\hsize]{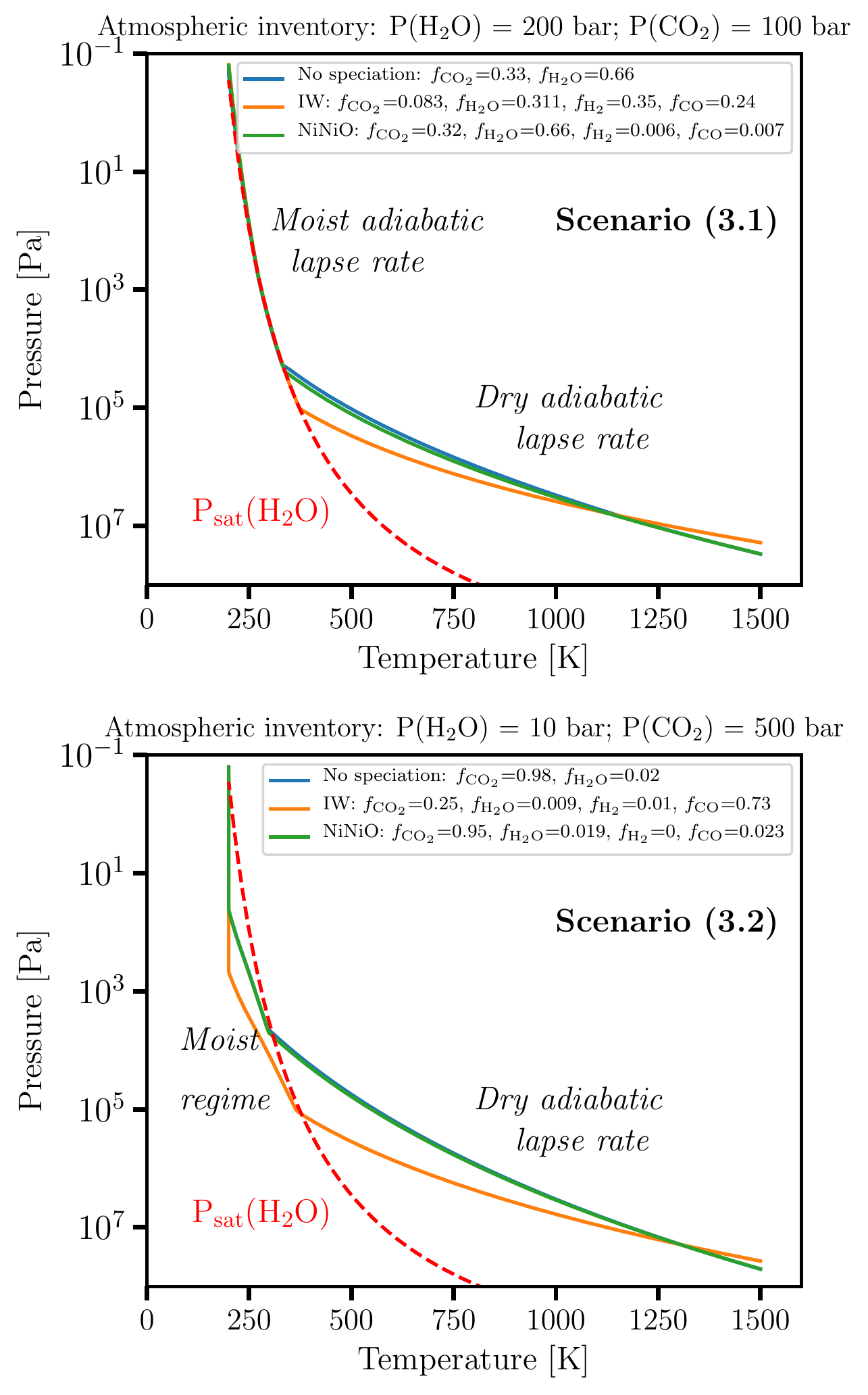}
 \centering
 \caption{Temperature profiles p-T for scenario 3.1 (upper panel) with an initially \ch{H2O}-dominated atmosphere  and scenario 3.2 (lower panel) with an initially \ch{CO2}-dominated atmosphere (Table~\ref{buffer}). The molar fraction of each species is shown in the legend. "No speciation" indicates that the values were input directly by the user and were not calculated by the C-O-H speciation model.} 
 
 
 \label{marcq1}
\end{figure}

Figure~\ref{marcq1} shows the temperature-pressure profiles obtained  for scenarios 3.1  (upper panel) and 3.2 (lower panel) with \ch{H2O} and \ch{CO2} dominated atmospheres of Table~\ref{buffer}, respectively. In the case of scenario 3.1 with its initially \ch{H2O}-dominated atmosphere (Fig.~\ref{marcq1} upper panel), for the reducing buffer IW case, the gradient of the dry adiabatic lapse rate is slightly reduced (orange line)
due to chemical speciation of H$_2$ and CO as compared to the "no speciation" case (blue line).  However, for an oxidizing buffer IW+4 (green line), the dry adiabatic lapse rate is similar to the case for "no speciation" because the volume mixing ratio of outgassed species is similar to the initial volatile outgassing. In Fig.~\ref{marcq1}, scenario 3.2 (lower panel), the chemical speciation of the initially assumed \ch{CO2}-dominated atmosphere (see Table~\ref{buffer}), leads to a CO-rich (IW buffer) and CO$_2$-rich (IW+4 buffer) atmosphere. Thus, the moist adiabatic lapse rate changes as compared to the upper panel results and shows a deviation from pure water vapor saturation curve (dotted red line). This is because of a low water vapor volume mixing ratio $f_{\rm H_2O}=0.02$ and a higher \ch{CO2} volume mixing ratio $f_{\rm CO_2}=0.98,$ leading to an unsaturated troposphere. Moreover, the vertical range of the wet adiabat regime (scenario 3.2) is smaller than in scenario 3.1 (upper panel), and the radiative regime ($T = 200 \rm  \,\ K$ isothermal profile) is evident at 40 Pa, for instance, for the green curve (lower panel) but  occurs near the top of the atmosphere, that is, around 0.1 Pa, for scenario 3.1.
The differing behavior in scenario 3.2 compared with scenario 3.1 arises mainly from the different chemical speciation (see panel legends), which affects the heat capacity and therefore the adiabatic gradients.


\subsection{Thermal spectral emission}
\begin{figure*}[!hbt]
 \includegraphics[width=1.0\textwidth]{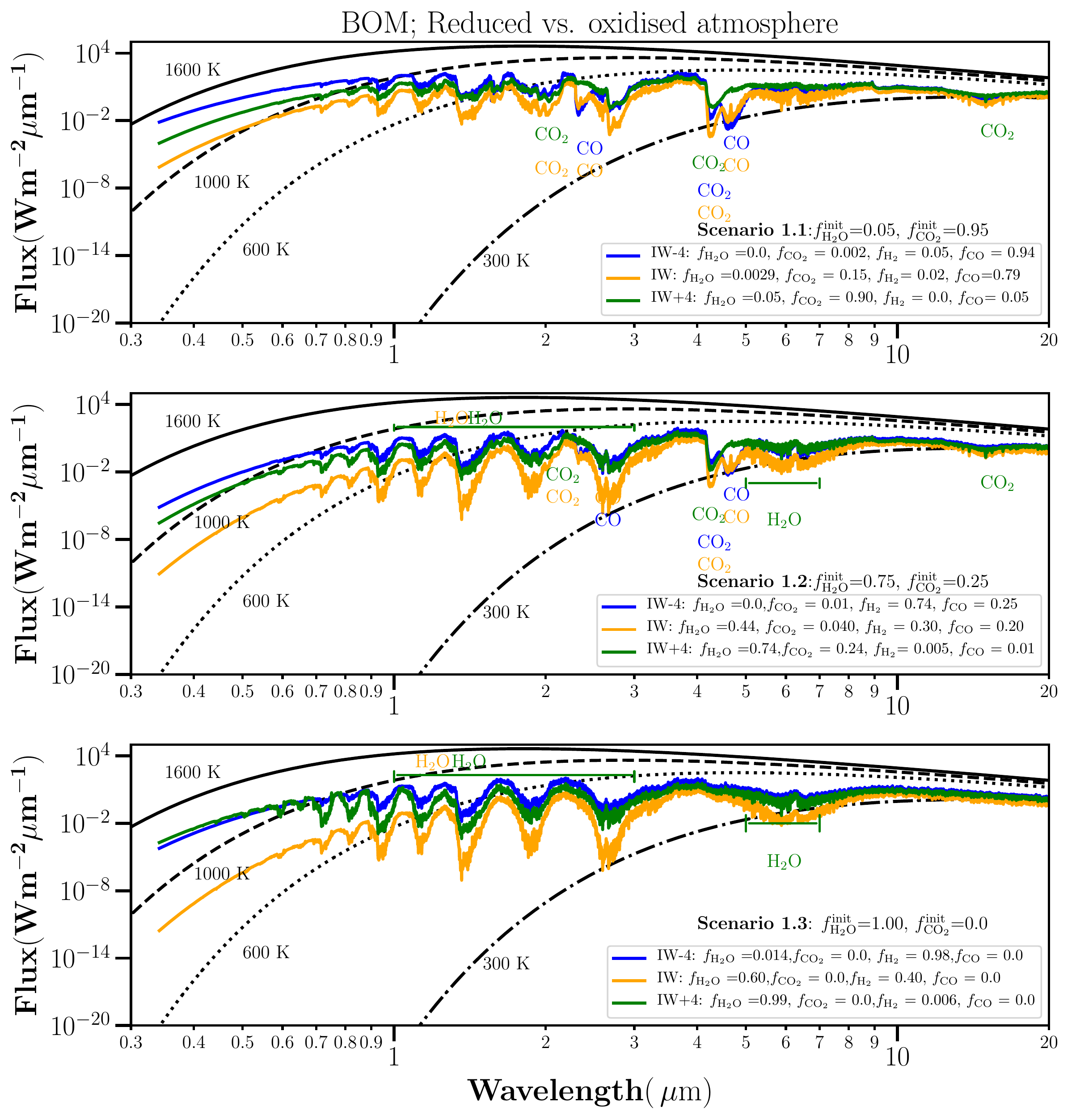}
 \centering
 \caption{Comparison of thermal emission spectra for a reduced vs. an oxidized mantle during the BOM presented for scenario 1.1 (upper panel), 1.2 (middle panel), and 1.3 (lower panel) of Table~\ref{buffer1}. The output of the volatile speciation model is shown in the legend. Values written above the legend show the initial input values to the speciation model. Reference blackbody curves for four different temperatures are plotted. Absorption features of key chemical species are indicated. The spectra shown here have been binned to a resolution of $\lambda/\Delta \lambda =1000$.
 }
 \label{BOM}
\end{figure*}

Using the temperature profiles as obtained in Fig.~\ref{redox_BOM}, Fig.~\ref{redox_EOM} and composition of various species for various scenarios of Table~\ref{buffer1} and Table~\ref{buffer2} as input to the line-by-line (lbl) radiative transfer code GARLIC, we obtain the thermal spectral emission from the atmosphere, also known as the outgoing longwave radiation (OLR).
Figure~\ref{BOM} shows the thermal emission spectra of scenario 1.1 (upper), 1.2 (middle), and 1.3 (lower) panel  of Table~\ref{buffer1} for the BOM. The spectral features that arise due to the presence of reduced species such as CO versus the oxidized species such as \ch{H2O} and \ch{CO2}
are easily distinguishable in the three panels. In the top panel of Fig.~\ref{BOM} (scenario 1.1), a prominent CO absorption feature is seen at 2.3 and 4.6 $\mu \rm m$ for the reduced buffer case IW-4 (blue curve), whereas  distinct \ch{CO2} features at 2.1, 4.3 and 15 $\mu \rm m$ are seen for the oxidized buffer case IW+4 (green curve). When the input \ch{H2O} outgassing value is increased (scenario 1.2), stronger \ch{H2O} features start to appear (orange and blue curves) along with the \ch{CO2} absorption features for the oxidized buffer (green curve). Finally, when the input \ch{H2O} outgassing is increased to 100\% (scenario 1.3), we obtain thermal spectra that are dominated by water vapor with distinct water features between 1 to 2.6  $\mu \rm m$ and at 6.2  $\mu \rm m$ for the oxidized case (green curve). The reduced atmosphere with IW-4 buffer produces an almost pure \ch{H2} atmosphere with a similar spectrum as \ch{H2O} but with a higher emitted flux. 

\begin{figure*}[!hbt]
 \includegraphics[width=1.0\textwidth]{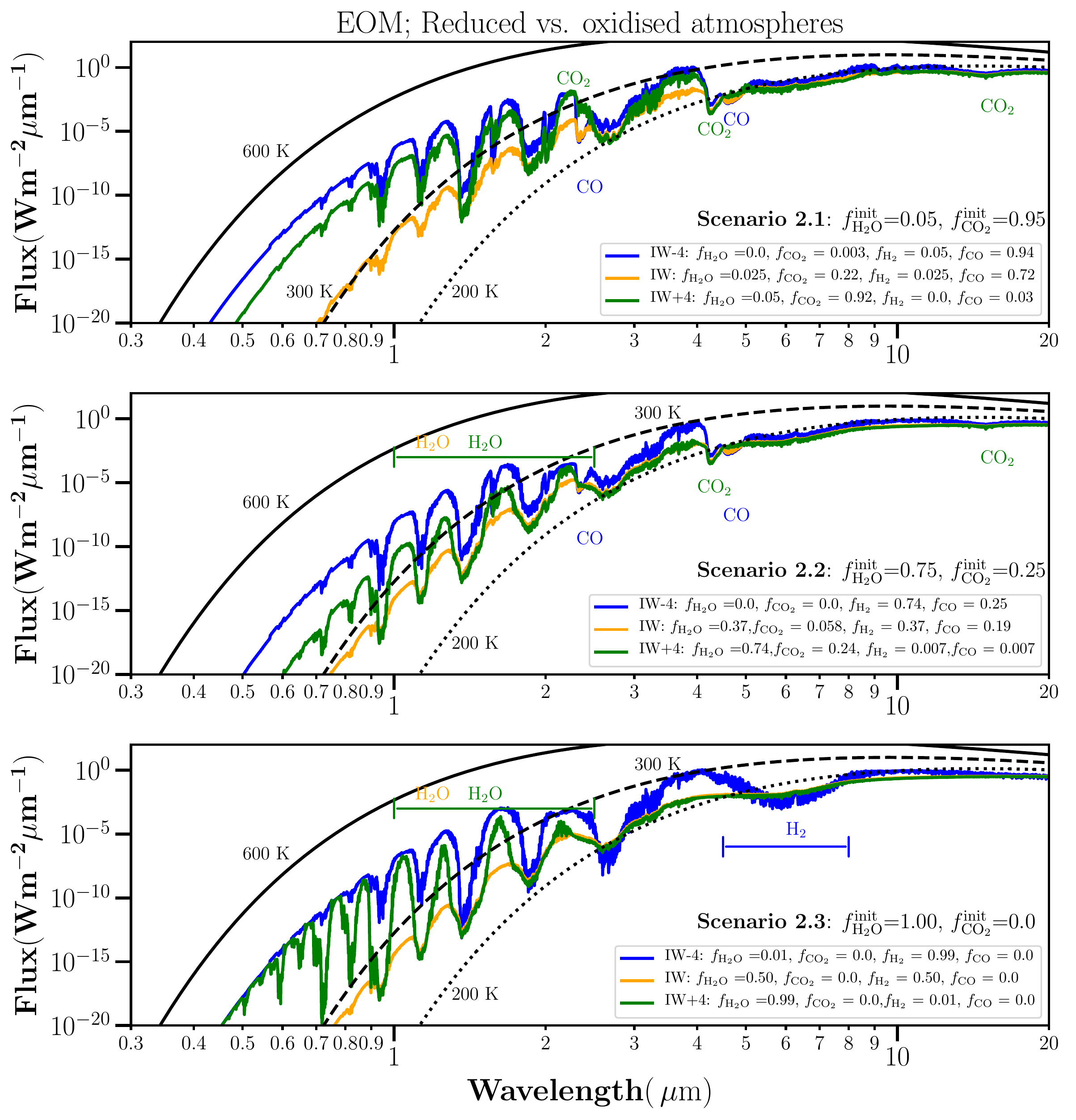}
 \centering
 \caption{Same as Fig.~\ref{BOM}, but for scenarios of Table~\ref{buffer2}. The blackbody curves for three different surface temperatures bracketing the emission spectra are plotted for reference.
 }
 \label{EOM}

\end{figure*}

Figure~\ref{EOM} shows the thermal emission spectra of scenario 2.1 (upper), scenario 2.2 (middle), and scenario 2.3 (lower) of Table~\ref{buffer2} for the EOM. One notable difference between Fig.~\ref{BOM} and  Fig.~\ref{EOM} is a significant reduction in the emitted flux especially at the smaller wavelengths, which is mainly related to the lower surface temperature (i.e., $1650 \rm \,\ K$) for the latter. The effect of pressure at the bottom of the atmosphere on the thermal spectra is also seen and is discussed in detail in Section~\ref{fug}. The effect of the large \ch{H2} continuum at 5-7 $\mu \rm m$ caused by self-collision of the \ch{H2} molecules could reveal molecular \ch{H2} as a main constituent of an atmosphere not known a priori.

\begin{figure*}[!hbt]
  \includegraphics[width=1\textwidth]{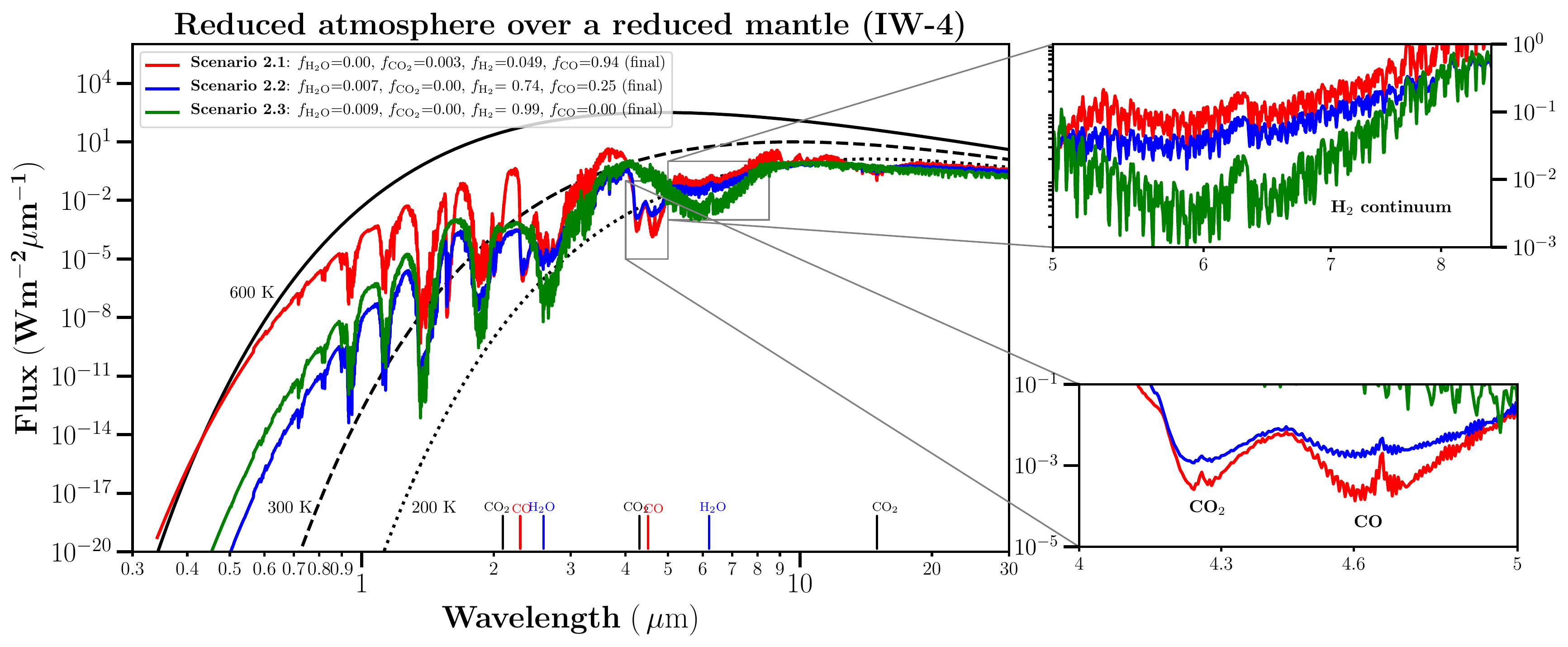}
 \centering
 \caption{Thermal spectral emission for a reducing atmosphere overlying a reduced mantle (IW-4) for scenarios 2.1, 2.2, and 2.3 of Table~\ref{buffer2}. The legend shows the mole fraction of the outgassed species obtained from the speciation model. Prominent features of absorbing species are seen at wavelengths marked in the figure. Reference blackbody curves for three different surface temperatures are plotted. The insets show the zoomed-in thermal spectra with the most prominent absorption features of \ch{CO2}, \ch{CO} and \ch{H2}. The spectra have been binned to a resolution of $\lambda/\Delta \lambda =1000$.}

 \label{IW-4}
\end{figure*}

Figure~\ref{IW-4} compares the thermal emission spectra for reduced atmospheres overlying a reducing buffer (IW-4) for the various scenarios 2.1, 2.2, and 2.3
of Table~\ref{buffer2}.  In this figure, the prominent absorption features are seen for CO at 2.3 and 4.6 $\rm \mu m$, while CO$_2$ stands out at 2.1, 4.3, and 15 $\rm \mu m$. A small amount of \ch{CO2} is always present even for this very reducing buffer calculated by the mass-equilibrium method (Sect.~\ref{s2}). However, the strength and width of \ch{CO2} absorption features are seen to decrease as the volume fraction of initial CO$_2$ decreases from 0.95 to 0.0 (see the zoomed 15 $\rm \mu m$ feature in the right inset of Fig.~\ref{IW-4}). Another notable feature is visible in scenario 2.3 with pure \ch{H2O} (dissolved in the melt), which results in a \ch{H2} -dominated atmosphere and displays a large continuum absorption in the 5-7 $\rm \mu m$ region caused by collision-induced absorption.

\begin{figure*}[!hbt]
 \includegraphics[width=1\textwidth]{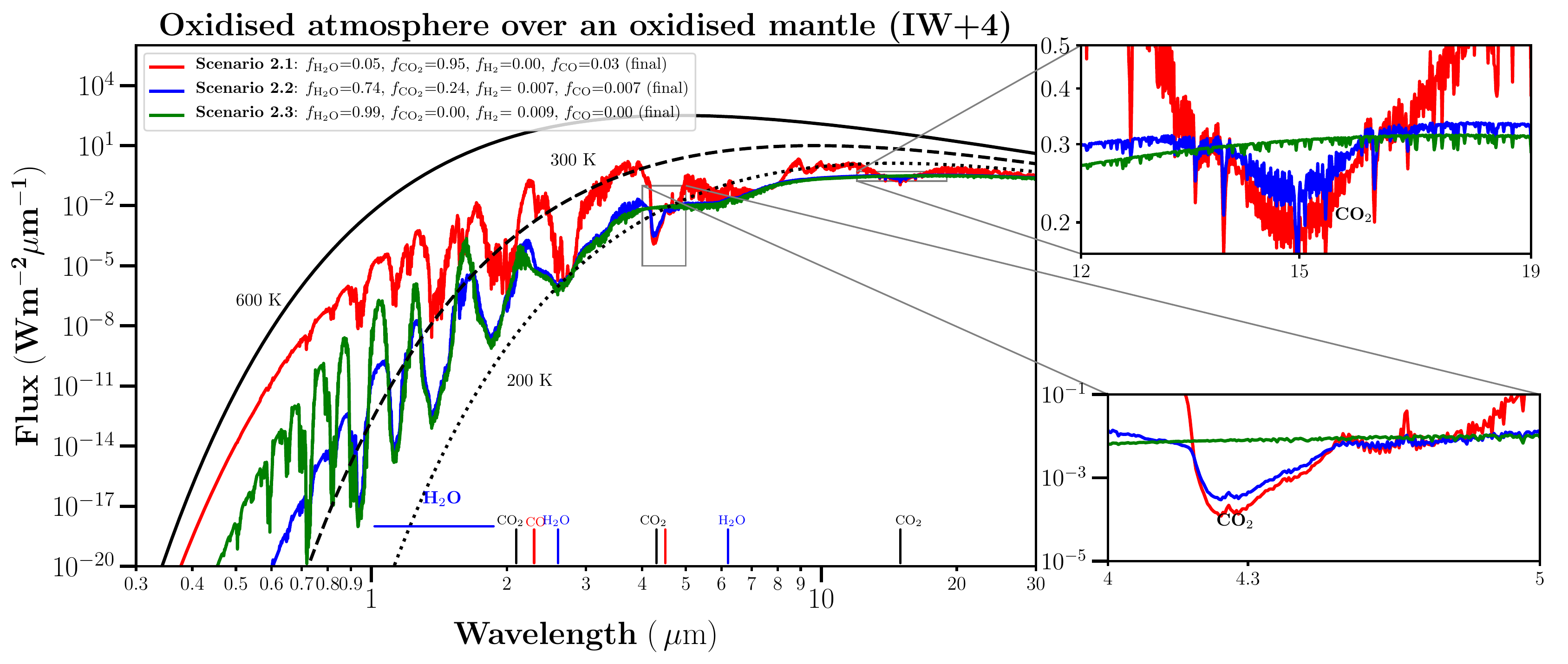}
 \centering
 \caption{Same as for Fig.~\ref{IW-4}, but for an oxidizing atmosphere overlying an oxidized mantle (IW+4). Notable features are the emergence of prominent \ch{H2O} features between $\rm 1-2 \; \mu m$ (scenario 2.3) and the disappearance of CO feature at $\rm 4.6 \; \mu m$ as compared to Fig.~\ref{IW-4} (scenario 2.1 and 2.2).  }
 \label{IW+4}
\end{figure*}

Figure~\ref{IW+4} is same as Fig.~\ref{IW-4}, but now for  the oxidizing atmosphere cases (IW+4) for scenarios 2.1, 2.2, and 2.3 of Table~\ref{buffer2}.
A reduction in the emitted flux is seen compared with the reduced atmosphere case (Fig.~\ref{IW-4}). This is because of enhanced absorption by higher concentrations of greenhouse gases such as \ch{CO2} and \ch{H2O} that are present in the oxidized atmosphere. Thus, as the initial \ch{H2O} volume mixing ratio (VMR) increases (from scenario 2.1 through 2.3), a further reduction in the emitted flux is obtained. The most prominent absorption features are seen for CO$_2$ at 2.1, 4.3, and 15 $\rm \mu m$ and \ch{H2O} at 1.0, 1.2, 1.4, 1.9, and 2.6 $\rm \mu m$, respectively.
As a result of the thick overlying steam atmosphere, the atmospheric window (8-10 $\mu \rm m$) becomes optically thick and the 6.2 $\rm \mu \rm m$ \ch{H2O} feature is not prominent (green curves in scenarios 2.2 and 2.3). The inset of this figure shows the zoomed-in 4.3 $\rm \mu m$ CO$_2$ feature. The CO feature at 4.6 $\rm \mu m$ is not seen here as compared to the reduced case in Fig.~\ref{IW-4}. The inset on the far right shows a zoomed-in view of the 15 $\rm \mu m$ CO$_2$ absorption band as a function of \ch{CO2} VMR and indicates the change in band depths when the input \ch{CO2} VMR is reduced from 0.95 to 0.25 (red to blue curve) and is then absent for the green curve with no \ch{CO2}. In summary, the main 
result is a diminishing of the CO band at 4.6 $\mu \rm m$ and the appearance of distinct \ch{H2O} bands from $\sim$ 1 - 2.6 $\mu \rm m$. The \ch{H2O} feature at 6.2 $\mu \rm m$ is less prominent because of the thick steam atmosphere.

\subsection{Dependence of OLR upon mantle fugacity}\label{fug}
\begin{figure*}[t]
  \begin{minipage}[c]{1\textwidth}
     \includegraphics[width=0.5\textwidth]{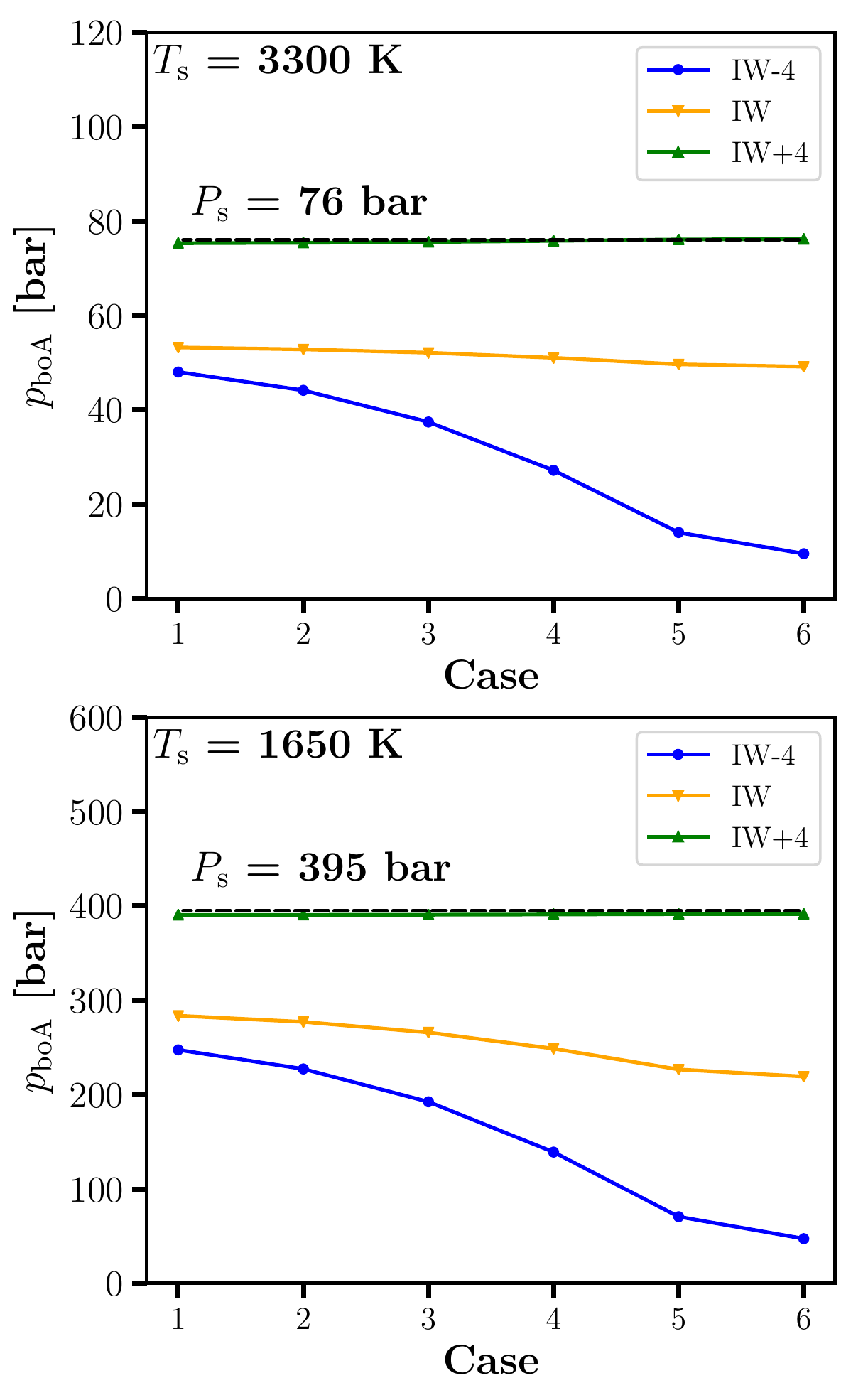}
 \includegraphics[width=0.5\textwidth]{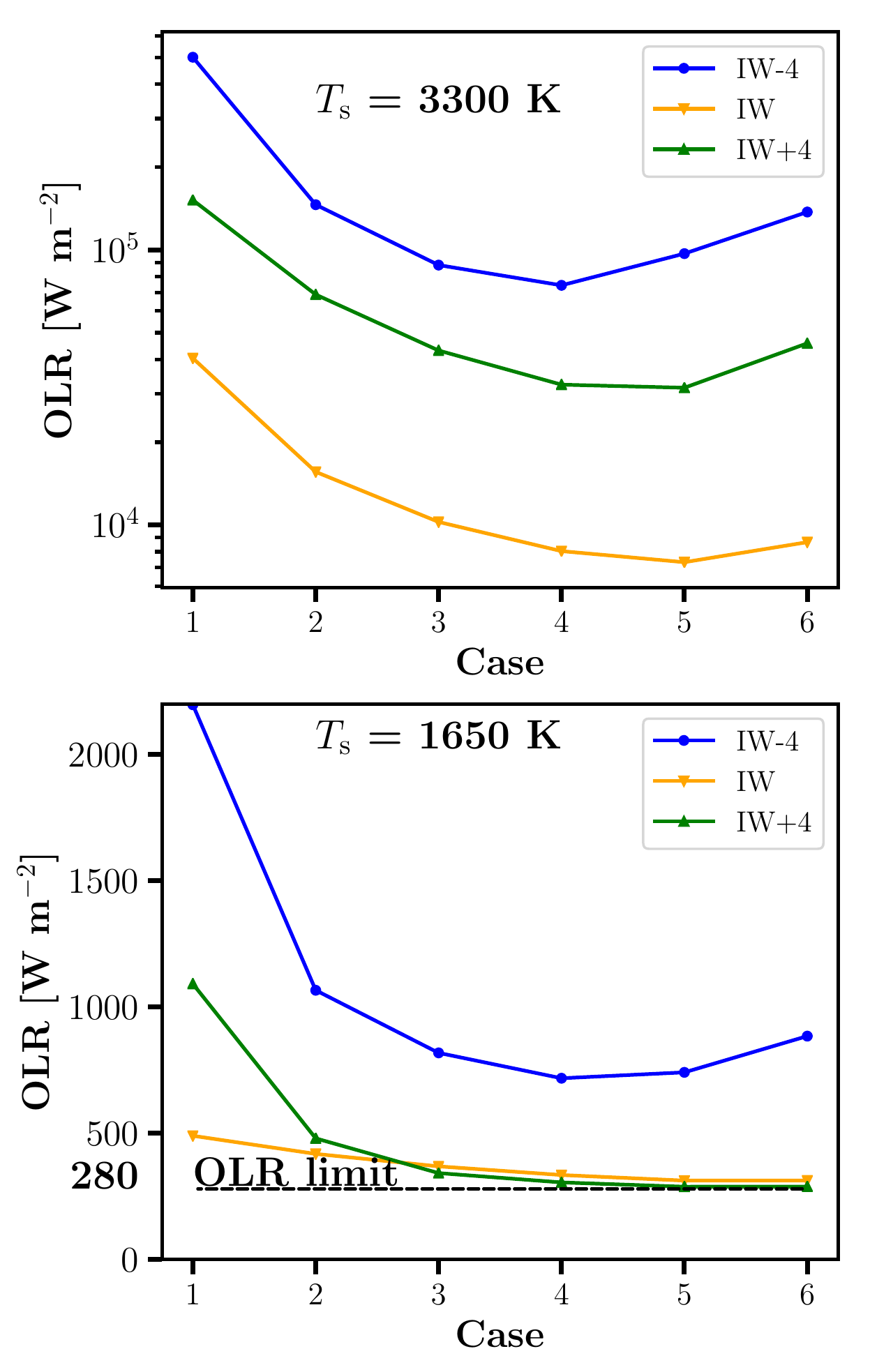}
  \end{minipage}
  \caption{{\it Left}: Pressure at the bottom of the atmosphere $p_{\rm boA}$ for various cases referring to different initial volatile abundances and three different mantle buffers: reduced (IW-4), IW, and oxidized (IW+4). The details of speciation for various cases on the x-axis are shown in Table~\ref{case1} ($T_{\rm s}=3300$ K) and Table~\ref{case2} ($T_{\rm s}=1650$ K) in the Appendix~\ref{App2}. The horizontal dashed black line indicates the total surface pressure of the initial outgassed volatile for the various cases. {\it Right}: Outgoing longwave radiation (OLR) for various cases corresponding to the atmospheric surface pressure shown in the left panels and for three different buffers.}
  \label{OLR}
\end{figure*}

The thermal emission spectra were then averaged over the entire wavelength range to obtain the outgoing longwave radiation (OLR) as shown in Figure~\ref{OLR} (right panels).
The variation in pressure at the bottom of atmosphere $p_{boA}$ (using Eq.~\ref{pboA}) for three different buffers is shown in Figure~\ref{OLR} (left panels). The details of the various cases for which we investigated the effect of speciation on the $p_{\rm boA}$ and subsequently on the OLR for the BOM and EOM are shown in Tables~\ref{case1} and \ref{case2} of Appendix~\ref{App2}, respectively.

  For the BOM (Fig.~\ref{OLR}; top right panel) with $T_{\rm s}=3300 \rm \; K$, the highest OLR is obtained for the most reduced atmosphere (blue curve; IW-4 buffer) because the absorption bands are weaker and the atmospheric pressures (top left panel) are lower than the IW buffer (orange curve), wherein a mix of reduced and oxidized species are present in the atmosphere (see Table~\ref{case1}) that contribute far more effectively to block radiation. It is interesting to note that even for the most oxidizing atmosphere (green curve; IW+4 buffer), less radiation is blocked by the atmosphere, which is mainly dominated by \ch{H2O} , than for the IW buffer. 

At lower temperatures (Fig.~\ref{OLR}; bottom right panel) with $T_{\rm s}= 1650 \rm \; K$ (EOM), the emitted radiation is far lower than in the high temperature case ($T_{\rm s}=3300$ K) because for the latter, the radiation is emitted in the infrared and visible wavelengths, which leads to a stronger outgoing longwave radiation (OLR). Moreover, for the low temperature case ($T_{\rm s}=1650 \rm \; K$), the OLR reaches a limit of $\sim$280 W/m$^2$ \citep{Nakajima1992,Goldblatt2013,Marcq2017} for water-dominated atmospheres (green curve; IW+4 buffer) at $\sim$1800 K \citep[also see ][]{Katyal2019}. The calculated OLR does not drop below the OLR limit because of the impeding effect of absorption of radiation by \ch{H2O} 
 even if present in moderate quantities. This is an important result that suggests that for a specific surface temperature, atmospheres with lower surface pressures ($\sim$4-200 bar) can lead to strong outgoing emission of radiation and therefore to a more effective cooling of the MO than for atmospheres with a high surface pressure (> 200 bar). Studying the effect of pressure on OLR in this manner is analogous to the effect of a threshold temperature as shown in \cite{Marcq2017}. We also refer to Figure 10 of \cite{Nasia2019}, where the OLR for pure steam atmospheres is plotted as a function of surface temperature and surface pressures. According to their Figure 10 and \cite{Katyal2019}, higher OLR values are obtained for atmospheres with lower surface pressure (for a similar surface temperature). This means that the outgoing radiation for lower pressure cases may reach the OLR limit, but at temperatures lower than $\sim 1800 \rm \,\ K$ as previously reported \citep{Kopparapu2013,Katyal2019}.





In Figure~\ref{OLR}, cases 1 and 2 are mainly dominated by \ch{CO2} and/or \ch{CO} (see Tables~\ref{case1} and \ref{case2}), which show only few absorption bands and thus display the
strongest OLR. Cases 3 and 4, which are composed of a mixture \ch{H2O}+\ch{CO2}+\ch{CO}+\ch{H2} (see Tables~\ref{case1} and \ref{case2}), result in a minimum OLR where the spectra shows absorption lines of almost all the species present.  Cases 5 and 6 are dominated by absorption through \ch{H2O} and/or \ch{H2}, which 
leads to a stronger OLR than the mixed-species cases 3 and 4, where absorption of all species strongly reduces the OLR for the high-temperature cases. For the low-temperature cases (lower right panel), the effect of the absorption by \ch{H2O}  dominates because the \ch{H2O} band and continuum absorption cover the entire infrared wavelength regime.

\subsection{Early \ch{H2} atmosphere  scenario}
While hydrogen is prone to escape to space via strong incoming extreme UV (EUV) during the Hadean (more details in Section~\ref{escape}), it is also an important gas in enhancing the greenhouse effect in the atmosphere. The efficiency of heating the atmosphere, however, depends upon the surface partial pressure of hydrogen. 

We explored a case similar to scenario 2.3 of Table~\ref{buffer1} and shown previously in Figure~\ref{IW-4} with a pure \ch{H2} atmosphere, but with a varying hydrogen surface pressure $p_{\ch{H2}}$ and a fixed surface temperature $T_{\rm s} = 2800 \, \rm K$. We obtained the thermal emission of radiation by varying the surface pressure $p_{\ch{H2}}$ in the range 0.6-185 bar, as shown in Figure~\ref{OLR_pressure} (upper panel) of Appendix~\ref{App3}.  We observe a large number of undulations around 2 $\mu \rm m$  in the obtained thermal spectra, especially at higher surface pressures. These undulations are caused by the effects of collision-induced absorption (CIA) by \ch{H2}.
The lower panel of Fig.~\ref{OLR_pressure} shows a decrease in the OLR with the (surface) partial pressure of \ch{H2} attributed to the higher absorption of radiation at higher pressures, as also shown in Fig.~\ref{OLR} (left panels). We therefore conclude that several bars of \ch{H2} in the atmosphere could significantly affect the climate in the cases studied, leading to additional greenhouse warming \citep[also see][]{Ramirez2014}. A similar trend in the OLR versus $p_{\ch{H2}}$ as shown here was suggested by \cite{Pierrehumbert2011}. 

\subsection{Transmission spectra}\label{trans}
We obtained the transmission spectra for the BOM and the EOM using the lbl radiative transfer code GARLIC and  the temperature profiles (Sect.~\ref{tp}) and concentration as an input. The resultant spectra are shown as the wavelength-dependent effective height of the atmosphere (see Eq.~\ref{height} in Figure~\ref{trans}). The spectra were binned to a resolution of $R = \lambda/\Delta \lambda= 1000$ over the wavelength range shown.

\begin{figure*}[bth]
 \includegraphics[width=1\textwidth]{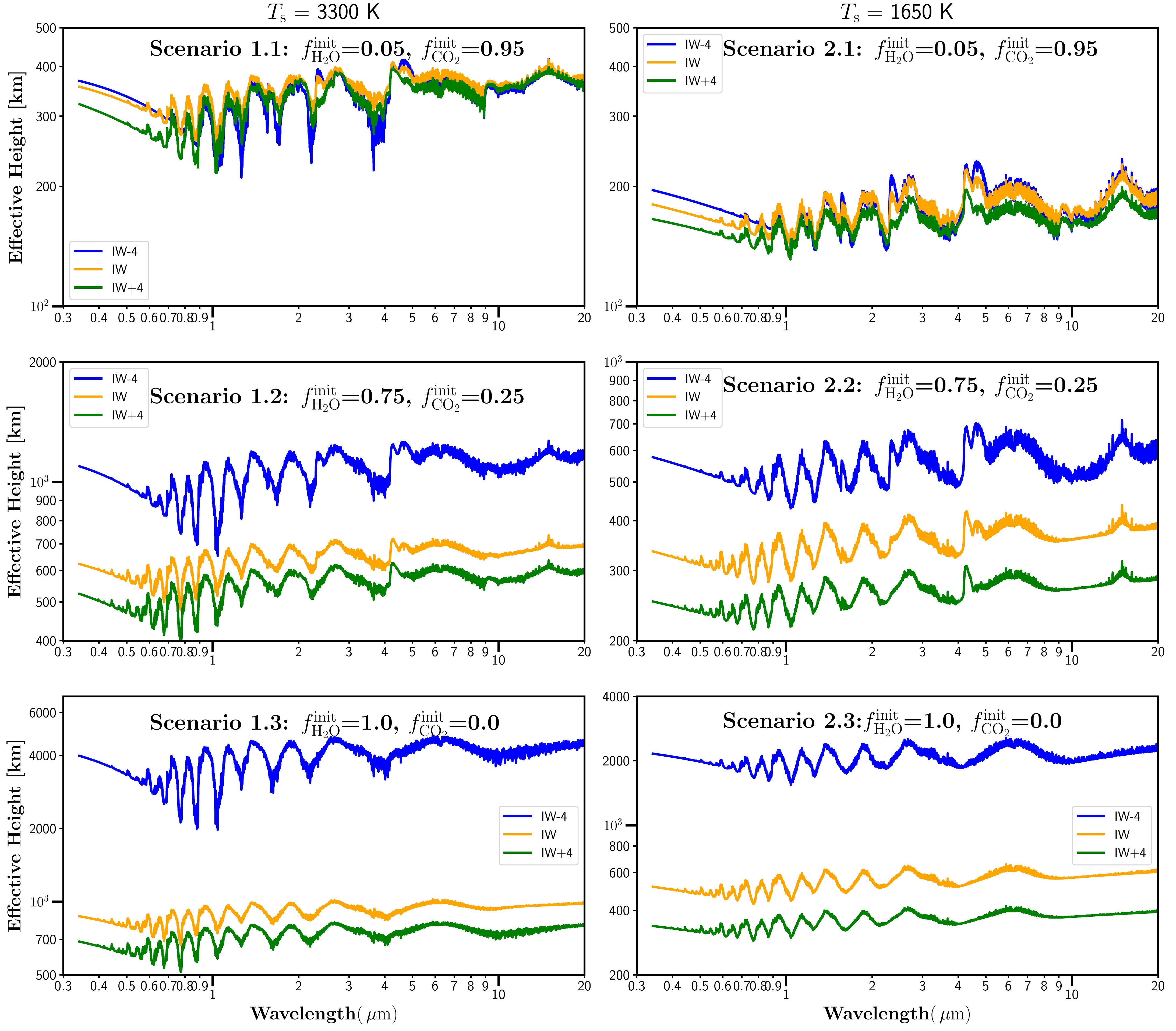}
 \centering
 \caption{Transmission spectra calculated by GARLIC shown as atmospheric effective height vs. wavelength for the BOM scenarios (left panels) and EOM scenarios (right panels) for three different redox buffers IW-4, IW, and IW+4. The mole fraction of final outgassed species in the atmosphere for various scenarios is the same as provided in Tables~\ref{buffer1} and \ref{buffer2}. The effective height of a reduced atmosphere (with low molecular weight) for the case of reduced mantle IW-4 (blue curve; scenarios 1.2, 1.3, 2.2, and 2.3) is lower than an atmosphere with larger molecular weight (blue curve; scenarios 1.1 and 2.1) and oxidizing atmospheres (green curve; all scenarios). For more details, see the text. The spectra have been binned to a resolution of $\lambda/\Delta \lambda=1000$.
  } 
 \label{trans} 
\end{figure*}

As shown in Fig.~\ref{trans}, the effective height of the atmosphere increases to one order of magnitude for the reduced atmospheres (IW-4) from scenarios 1.1 to 1.3 (and 2.1 to 2.3) for both BOM (left panel) and EOM (right panel) scenarios. This is related to the increasing atmospheric scale height for the lighter \ch{H2} -dominated atmospheres. Similarly, for the oxidized atmospheres (IW+4), the effective height increases from a dense and thick \ch{CO2} -dominated atmosphere (scenarios 1.1 and 2.1) to lighter \ch{H2O} -dominated atmospheres (scenarios 1.2, 1.3, 2.2, and 2.3).
In the visible-optical range, the slope of the so-called Rayleigh-scattering extinction feature is somewhat steeper for the reducing atmosphere (blue line) because of the different scale heights. For scenarios 1.3 and 2.3, \ch{H2O} absorption features are seen ubiquitously from roughly 0.9 to 2.6 $\rm \mu m$. 
For a high \ch{CO2} initial volatile outgassing (scenario 1.1 and 2.1), CO absorption features are seen at 2.3, 4.6 $\rm \mu m$ for IW-4 buffer, and \ch{CO2} absorption features are seen  as shown for 2.1, 4.3, and 15 $\rm \mu m$ for the IW+4 buffer. 
For the EOM scenarios (right panels), the cooler surface temperature (1650 K) leads to a decrease in scale height ($H=kT/mg$) compared with the BOM (left panels) scenarios with higher surface temperature (3300 K).


A decrease in the depth of the absorption bands for CO (IW-4) and \ch{CO2} (IW+4) is seen at 4.3 and $\rm 15 \; \mu \rm m$,  respectively, which is related to a decrease in the input \ch{CO2} from scenarios 1.1 to 1.2 at $T_{\rm s}=1650 \; \rm K$ (right panel). A similar trend is seen for scenarios at $T_{\rm s}=1650 \; \rm K$ (right panel). For the oxidizing atmospheres (IW+4) in scenarios 1.3 and 2.3, the most prominent \ch{H2O} features between $\rm 1-2 \; \mu \rm m$ are seen, but the 6.2 $\rm \mu m$ \ch{H2O} feature is suppressed.  This is likely related to the overlying optically thick atmosphere of \ch{H2O} (blanketing effect) with a very high surface pressure ($\sim 395 \rm bar$), wherein the radiation will be dominated by the emission from the top of the atmosphere. This hinders chemical species in the lower atmosphere from being detected in the upper regions sampled by transmission spectroscopy. It is therefore  difficult to probe dense atmospheres with high optical depths.

\subsection{Atmospheric evolution and escape over the MO period} \label{escape}
Table~\ref{evolution} shows the range of surface partial pressure of \ch{H2} for a highly reduced (IW-4) and an oxidized mantle (IW+4) over a period of MO solidification $\sim 1 \; \rm Myr$. For the reducing-buffer case (IW-4), over the period of MO, the surface pressure of atmospheric hydrogen increases from 2.3 to 102 bar, while for the oxidized case, it increases from 0.02 to 2.7 bar. To investigate the evolution of \ch{H2} along with other species in the atmosphere, we split the MO period into ten time steps where the change in species abundance from scenario 1.1 (BOM) to scenario 2.2 (EOM) in Table~\ref{evolution} is set to change with a linear time step of 0.1 Myr (only shown for \ch{H2} in the table). The resulting changes in species abundances across the MO can be attributed to outgassing and chemical speciation. As a first step, the input data for the various scenarios were taken from \cite{Nasia2019,Katyal2019}, but in a future study, we plan to perform time-dependent, coupled interior-atmosphere evolutionary scenarios over the
magma ocean phase applying a newly developed climate-chemistry atmospheric model \citep{Markus2020,Fabian2020}.

\begin{table*}[!hbt]
    \centering
    \caption{Hydrogen abundance in the atmosphere from the beginning (scenario 1.1) to the end (scenario 2.2) of the magma ocean for a reduced (IW-4) and oxidized (IW-4) mantle. }
    \large
\begin{tabular}{||c||c|c|c||c|c|c||}
\hline
\hline
Buffer & \multicolumn{3}{c||}{t = 1 yr (scenario 1.1)} & %
    \multicolumn{3}{c|}{t = 1 Myr (scenario 2.2)}  \\
\cline{2-7}
\cline{2-7}

 & $p_{boA}$(bar) & $p_{\ch{H2}}$ (bar) & $f_{\ch{H2}}$&$p_{boA}$(bar)& $p_{\ch{H2}}$ (bar)&$f_{\ch{H2}}$ \\
\hline
IW-4 &48 &2.3 &0.05 &139 &102.8& 0.74\\
\hline
IW+4 & 75.3& 0.02& 0.0003 &391 &2.7 &0.007 \\
\hline
\end{tabular}

\label{evolution}
\end{table*}

We now consider the evolution of atmospheric species over the MO period. The mole fraction $f$ of the resultant outgassed species $i$ is translated into the mass fraction via $w_i= f_i \; M_i /\bar M$, where $M_i$ is the molecular weight of the species $i$ and $\bar M$ is the mean molar mass of the atmosphere. The atmospheric mass of the individual species is obtained by multiplying the calculated mass fraction $w_i$ with the total mass of the atmosphere $M = P \mbox{*} A/g$. 

\begin{figure*}[!hbt]
\centering
  \begin{minipage}[c]{0.50\textwidth}
  \hspace{-20pt}
    \includegraphics[trim=1cm 1.5cm 0 10cm ,clip,width=1.4\textwidth]{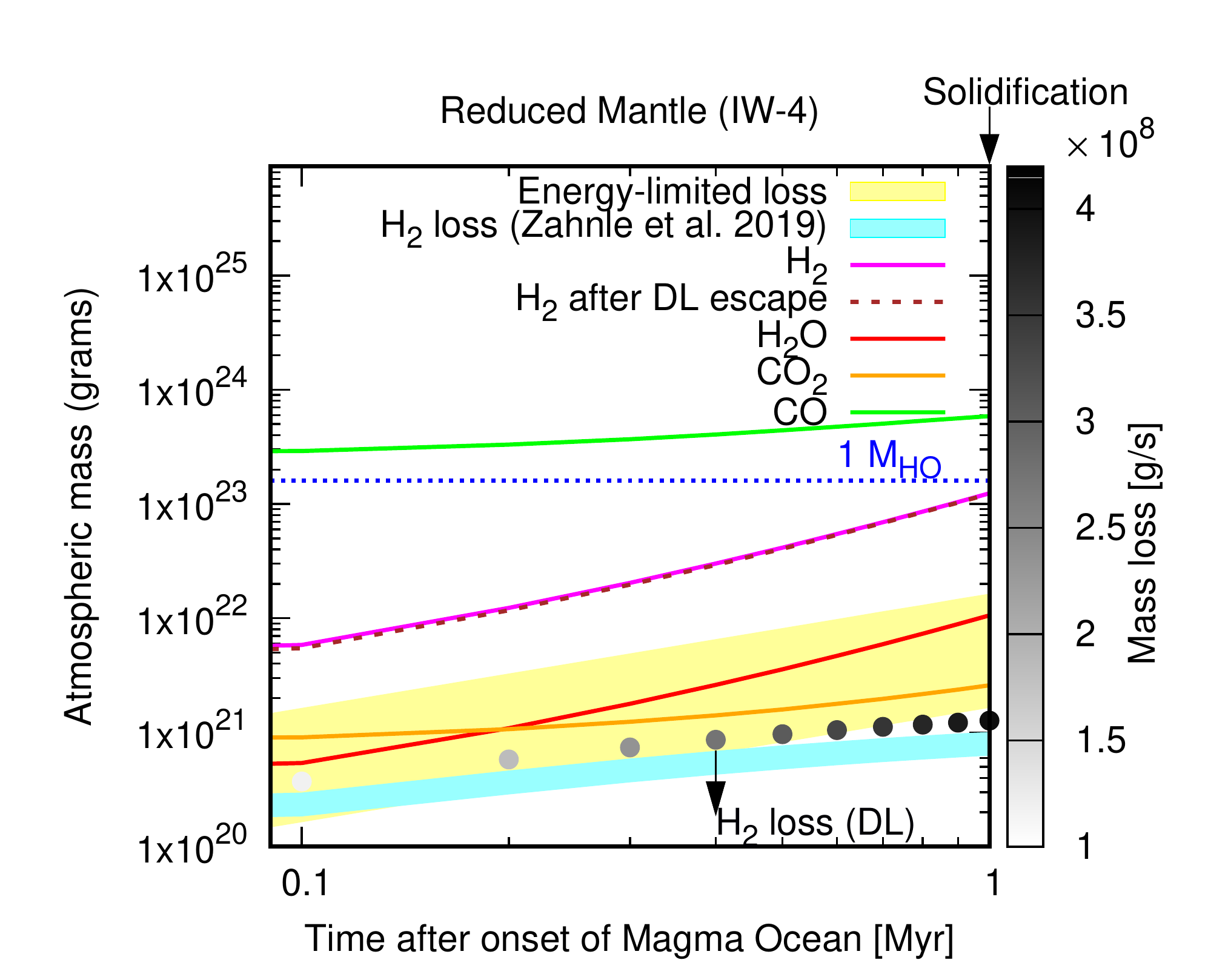}

  \end{minipage}
  
  \begin{minipage}[c]{0.5\textwidth}
    
 \hspace{-20pt}
 \includegraphics[trim=1cm 1cm 0 11cm ,clip,width=1.4\textwidth]{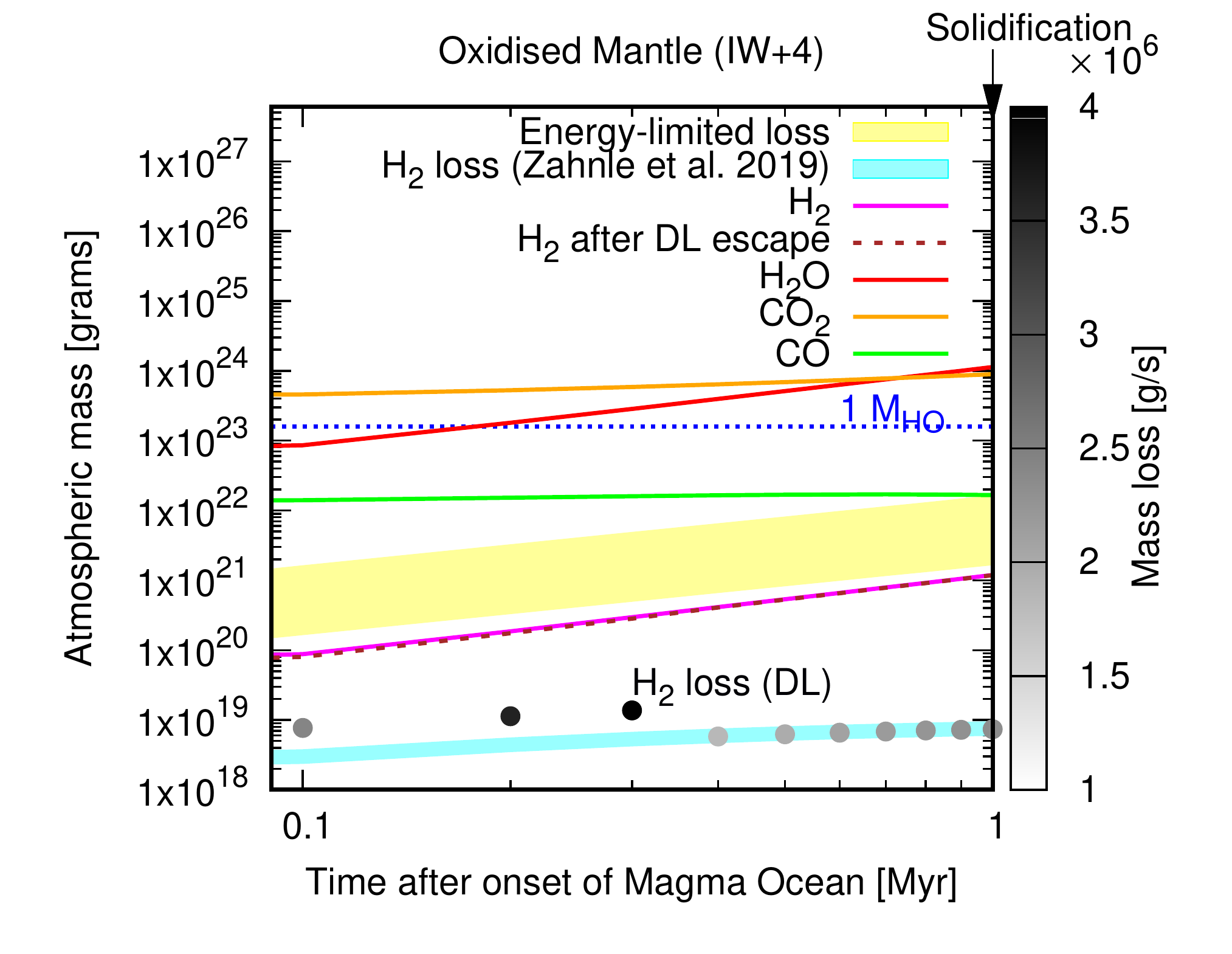}
  \end{minipage}
  \caption{Evolution/escape of atmospheric species in terms of the atmospheric mass (left axis) for the duration of the magma ocean (x-axis) for a reduced mantle case (upper panel) and an oxidized mantle case (lower panel). The colored solid lines refer to atmospheric masses integrated over the whole planet that change due to outgassing and chemical speciation (see text).  The dotted brown line shows the residual mass of \ch{H2} in the atmosphere after accounting for outgassing and DL escape. 
    The shaded region in cyan shows the mass loss of \ch{H2} in the range for $S=10$ to $S=100$ (XUV relative to the modern value) using the \cite{Zahnle2019} best-fit formalism (see text). The shaded region in yellow shows the energy-limited loss of \ch{H2} for the lower range $S=10$ to upper range $S=100$. The dotted blue horizontal line shows the hydrogen content of one (modern) Earth ocean (M$_{\rm HO}=1.6 \times 10^{23}\rm \,\ g$). The filled circles indicate the escaping mass of \ch{H2} obtained using DL. The colors indicate the respective mass-loss rates (right axis).}
  \label{loss}
  \end{figure*}

Figure~\ref{loss} shows the resulting atmospheric mass evolution of a reducing atmosphere lying above a reduced mantle, that is, IW-4 (upper panel) and an oxidizing atmosphere lying above an oxidized mantle, that is, IW+4 (lower panel) during the MO period. The circles show the mass of the \ch{H2} escaping via diffusion-limited escape according to Eq.~\eqref{dm} as described in Section~\ref{escape_basic}, where the color of the circles indicates the total mass-loss rate (g/s). The mass-loss rate obtained from energy-limited escape formalism as in Eq.~\eqref{el1} is obtained to be $5.2 \times 10^{7} \; \rm g/s$ for $S=10$ and $5.2 \times 10^{8} \; \rm g/s$ for $S=100$.  As described in Sect.~\ref{escape_basic}, ($R_p + H_e) \sim R_p$ for this scenario (see Table~\ref{evolution}). The mass loss of \ch{H2} (in grams) is then calculated and shown as the yellow shaded region within these $S$ ranges in Fig.~\ref{loss} (both upper and lower panels).  In general, the atmospheric \ch{H2} mass loss (in grams) is seen to increase for both the escape processes because the \ch{H2} abundance increases throughout the MO evolution timescale. The residual \ch{H2} in the atmosphere after accounting for escape is seen to overlie  the total \ch{H2} content of the atmosphere, implying that a significant amount of \ch{H2} remains in the atmosphere. We also show a shaded region obtained from the parameterized Eq.\eqref{zahnlebestfit} of \cite{Zahnle2019}, where the lower and upper boundaries  of the shaded region indicate $S = 10$ and $S=100$, respectively, and $f_{\ch{H2}}$ values are taken over the evolution timescale (see Table~\ref{evolution}). 

For the oxidized-mantle (IW+4) case (Fig.~\ref{loss} lower panel), the \ch{H2} abundance is very low (see Table~\ref{evolution}), causing this to be the bottleneck for \ch{H2} to diffuse through heavy \ch{CO2} or \ch{H2O} up to the homopause.  Moreover, the plentiful amount of XUV energy available here (with a high loss rate) can in fact lead to the complete removal of hydrogen.
For the oxidizing atmosphere lying above an oxidized mantle, that is, IW+4, escape of \ch{H2} is therefore  not energy limited but diffusion limited. Finally, we conclude that for both the redox states of the mantle, the outgassing of \ch{H2} into the atmosphere dominates the escape of \ch{H2} \citep{Kuramoto2013,Ramirez2014}.
 

\begin{figure}[!hbt]
 \includegraphics[trim=1cm 1cm 0 13cm ,clip,width=0.56\textwidth]{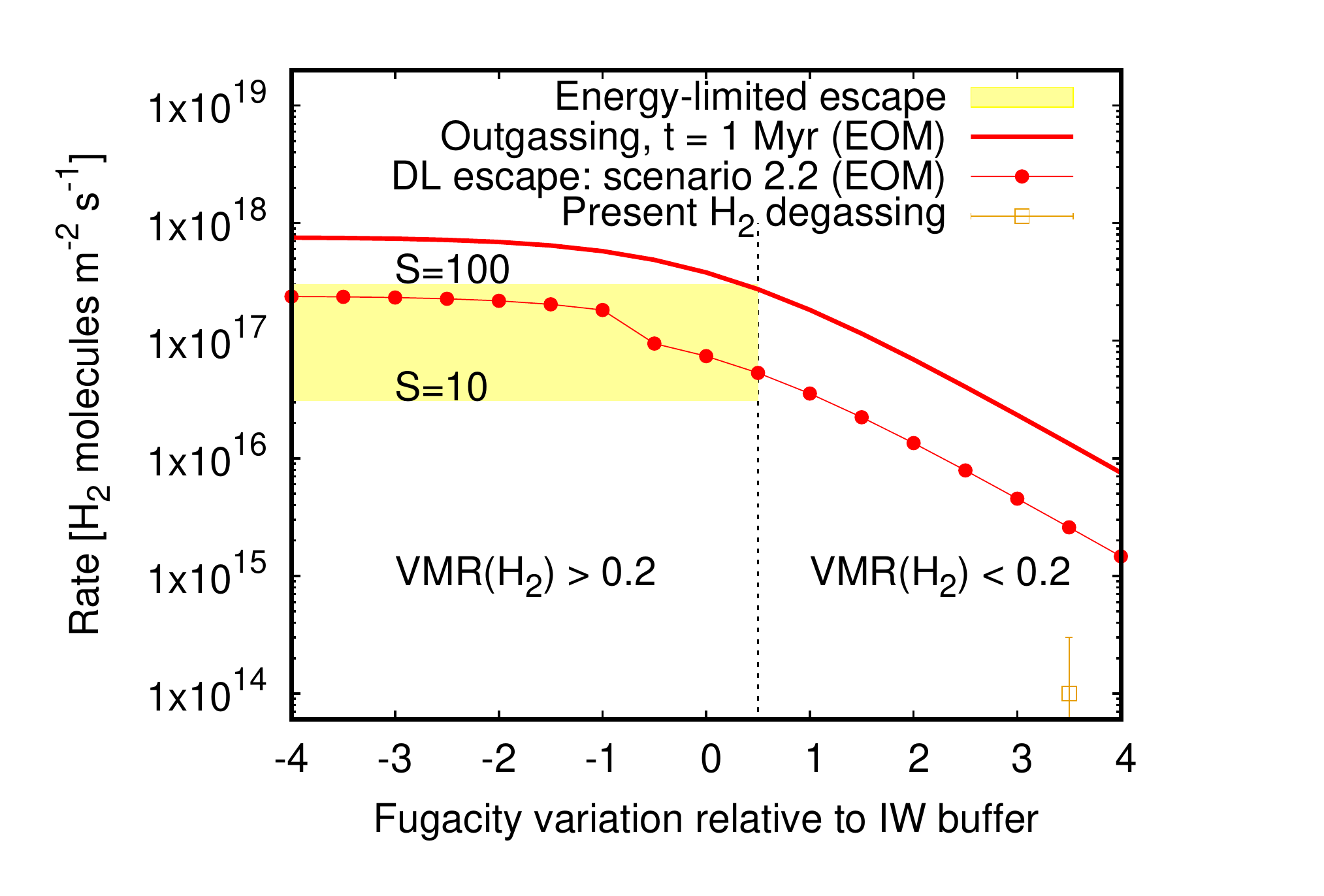}
 \caption{Hydrogen escape and outgassing rates as a function of redox state of the mantle for scenario 2.2  (EOM). The solid red line shows the \ch{H2} outgassing rate at the end of magma ocean. The closed red circles show the diffusion-limited escape of \ch{H2} for the EOM. The shaded region in yellow shows the loss rate due to energy-limited  escape within the range S=10 (lower boundary) to S=100 (upper boundary) i.e. XUV relative to the modern value. The vertical dotted line marks the region separating a high \ch{H2} VMR to low \ch{H2} VMR and illustrates the most efficient escape process occurring. The present Earth \ch{H2} degassing rate with an uncertainty range \citep{Ramirez2014} is marked for reference.}
 \label{DL}
\end{figure}

As an illustration, Figure~\ref{DL} compares the escape rates for scenario 2.2 (EOM) calculated for \ch{H2} with our estimated H$_2$ outgassing rate obtained using Equation~\eqref{molecules} as described in Section~\ref{s2}.

The outgassing rates obtained by us during the magma ocean period in the Hadean (3.8-4.0 Gyr before present) for a variable redox state of the mantle are higher than in the Archean (2-4.0 Ga), that is,  $9 \times 10^{14} \rm \,\ molecules \,\  m^{-2} s^{-1}$ \citep{Tian2005a,Kuramoto2013}. First, one of our central results is that outgassing (red solid line)  proceeds faster than  diffusion-limited escape (closed red circles) by a factor of x10 and energy-limited escape depending upon the uncertainty in the XUV radiation as evident from the $S$ values displayed in the figure. Second, for an atmosphere with a high hydrogen abundance, that is, \rm VMR(\ch{H2}) > 0.2  (-4 < IW $\le$ 0.5), the energy-limited escape is more effective, whereas for an atmosphere with lower hydrogen abundances, that is, $\rm VMR(\ch{H2}) < 0.2 \; (0.5 <IW <4)$, the diffusion-limited escape is more effective, as shown in Fig.~\ref{DL}.

\section{Discussion} \label{s4}
We have studied the effect of the redox state of the mantle and volatile outgassing  of major outgassed species \ch{H2O} and \ch{CO2} from the interior and their interaction with the melt resulting in outgassing of newer species at the surface. The effect on the atmospheric profiles, thermal infrared emission, transmission, and escape of \ch{H2} is discussed below. 


\subsection{Mantle redox and atmospheric temperature}\label{redox}


Our study has suggested an interesting coupling between interior outgassing, volatile speciation due to the mantle redox state, atmospheric composition, and atmospheric temperature that affects atmospheric extent, mixing, and the location of the dry and moist adiabat. During the BOM, more oxidized atmospheres switch from dry to wet adiabat at lower pressures (Fig.~\ref{redox_BOM}, green lines) than reduced atmospheres due to differences in their mean molecular weight and heat capacities, which leads to a lower scale height. The pressure at which the moist adiabat is reached affects the onset of saturation, thereby affecting the microphysical processes such as condensation and ultimately ocean formation. Detailed microphysical processes are not included in our study and need further investigation with more detailed models. On the other hand, the more reduced atmospheres have larger scale heights, and the switch to the moist adiabat lies at higher altitudes (Fig.~\ref{redox_BOM}, blue curve). Precipitation initiated at high altitudes has to sediment over a greater distance to rain out, which could favor evaporation in the lower, warmer atmospheric layers, and which might therefore be less efficient at condensing to form the oceans. For the EOM case (Figure~\ref{redox_EOM}), although the atmosphere cools (from 3300K to 1650K at the surface) and the surface pressure increases by a factor of 5, the temperature structures for the EOM are similar to those for the BOM (Figure~\ref{redox_BOM}). This is because the pressure increase (which increases atmospheric thickness) is offset by the cooling effect (which decreases atmospheric thickness). In summary, the above processes likely affected the evolution of climate on the early Earth by affecting key processes such as cooling to space and greenhouse heating, but also the onset of condensation and accordingly, the formation of terrestrial ocean.

\subsection{Thermal emission and transmission spectra}
Our results suggest that the thermal emission of radiation is affected by key parameters such as surface temperature, surface pressure, mantle redox state, and initial volatile content 
 of the two main species \ch{H2O} and \ch{CO2} that are outgassed during the magma ocean phase. The most prominent absorption features of oxidized versus reduced species in the atmosphere for the three different mantle oxidation states 
 IW-4, IW, and IW+4 are shown in Figs.~\ref{BOM} and \ref{EOM}. For the reducing mantle cases, the volatile speciation model would result in a mixture of both oxidized and reduced species in the atmosphere depending upon the initial volatile contents $f^{\rm init}_{\ch{H2O}}$ and $f^{\rm  init}_{\ch{CO2}}$ (see Tables~\ref{buffer1},  \ref{buffer2}, \ref{case1}, and \ref{case2}). As shown in Fig.~\ref{OLR} (right panels), the OLR for the most reduced atmosphere case overlying a reduced mantle (IW-4) consists of \ch{H2}, CO or CO+\ch{H2} (depending upon the case) and results in the highest OLR because these molecules have only a few absorption bands. On the other hand, we obtain lower OLR for other moderately reduced buffer, that is, IW, resulting in a mixed atmosphere of \ch{CO2}+\ch{CO}+\ch{H2}+\ch{H2O} with the largest number of absorption bands covering the entire spectral range. Finally, the lowest OLR is obtained for an oxidizing atmosphere overlying an oxidizing buffer (IW+4), resulting in \ch{CO2}, \ch{H2O} or \ch{H2O}+\ch{CO2} atmospheres (depending upon the case). 
 
 As a result, a planet with a reduced mantle buffer (IW-4) and low water content (cases 1 and 2) is expected to cool faster than a  planet with a high water content in the atmosphere (cases 4, 5, and 6). For a moderately reduced mantle buffer (IW), the planet will cool down slowly as the OLR is the lowest for it. For a planet with an  oxidized mantle buffer (IW+4), the cooling timescales will be slower than those of the IW-4, but faster than the IW case. 
 
 The initial outgassed water content along with the surface temperature and the redox state of the mantle are important factors driving planetary cooling. Figure~\ref{OLR} shows for lower temperature ($T_{\rm s}=1650$ K) and oxidized mantle (IW+4 buffer) that the OLR limit $\sim$280 W\; m$^{-2}$ \citep{Goldblatt2013,Marcq2017} occurs only for \ch{H2O} -rich atmospheres (cases 5 and 6). However, this is not the case for a high temperature ($T_{\rm s}=3300$ K) and an oxidized mantle (IW+4) because at such high temperatures, the emission occurs in visible wavelengths along with infrared, and \ch{H2O} is not able to block the radiation in the visible as effectively as in infrared. Radiation is thus not able to reach the OLR limit.
 
 Oxidized atmospheres containing heavier species such as \ch{CO2} result in a dense atmosphere with small effective heights for transmission spectra (Fig.~\ref{trans}, scenarios 1.1 and 2.1) as compared to a reduced and lighter atmosphere with \ch{H2}, resulting in extended atmospheres with a large effective scale height, as shown in scenarios 1.3 and 2.3 of  Fig.~\ref{trans}. The slope of the Rayleigh-scattering feature in the ultraviolet and visible wavelength range could provide information on the bulk atmospheric composition. However, when aerosols, thick hazes, or clouds are considered, this slope changes considerably because of the change in the wavelength dependence of the scattering (see Eq.~\ref{cloud}), which changes the atmospheric optical depth and effective height of the atmosphere \cite[also see][]{Fabian2020}.

\subsection{Outgassing and escape: loss timescales}
Our results suggest very different regimes in terms of outgassing of various species and escape of \ch{H2}  depending on the mantle redox state during the MO period. For the scenarios that consider the reducing and oxidizing state of the mantle, the outgassing rate of \ch{H2} is greater than the escape rate, and there is enough interior outgassing of \ch{H2} to form an atmosphere. A similar result was suggested by \cite{Ramirez2014} for early Earth and Mars.  

Because the VMR of \ch{H2} is high (e.g., in a reducing atmosphere), the escape of \ch{H2} most likely occurs by energy-driven, that is, energy-limited, escape. On the other hand, if less hydrogen is present (e.g., in an oxidizing atmosphere), the diffusion of hydrogen through a heavy gas to reach the upper atmospheric regions and escape is limited. Therefore, the atmosphere would probably enter a DL  configuration. Thus, depending upon the abundance of \ch{H2} in the atmosphere, our results suggest a switch from energy-limited (EL) escape (high VMR) of \ch{H2} to DL escape (low VMR), thus complying with \cite{Zahnle2019}.

For a reduced early atmosphere with 102 bar of \ch{H2} by the end of MO, we estimate the timescale of \ch{H2} atmosphere removal to be $\sim$10 Myr using both DL and EL ($S=10$) escape. On the other hand, because the \ch{H2} mass-loss rate and the surface partial pressure of surface hydrogen are lower (2.7 bar) for an oxidized atmosphere, the mass-loss timescale is estimated to be $\sim$16 Myr
assuming  DL escape. Similar mass-loss timescales ($\sim$10$^{6}$-10$^{7}$ yr) for surface pressures of 10 to 100 bar are obtained by hydrodynamic escape modeling of \cite{Pahlevan2019}. The results for total hydrogen mass-loss obtained via two different parameterized escape models (this work) and hydrodynamic model \citep{Pahlevan2019} are therefore approximately similar. Moreover, as we showed in Fig.~\ref{loss}, the mass loss obtained using the energy-limited (hydrodynamic) approach following \cite{Zahnle2019} (cyan shaded region)  approximates the diffusion-limited mass loss at higher levels of irradiation, $S=100$ \citep[also see][]{Lammer2018}. Furthermore, it is worth stating that several works \citep[e.g.,][]{Lammer2012} have noted the potential importance of EUV in driving energy-limited escape on early Earth and speculated that this quantity is not known to within a factor 100 or more times the modern-day value depending upon whether the early Sun was a fast or slow rotator. 


\subsection{Clouds}

\label{clouds1}

Clouds could be rather common features in extrasolar atmospheres, and they could have a potentially strong
effect upon the atmospheric spectra, climate, and photochemistry. Spectral features can be significantly reduced depending upon
the extent (layer location and thickness) and the microphysical properties (size, shape, number density distribution,
and refractive index) of the clouds occurring over a wavelength range that is sensitive to the cloud diameter, for example.
The general effect of clouds upon atmospheric spectra and retrieval by the James Webb Space Telescope (JWST) was discussed by \cite{Mai2019} and \cite{Fauchez2019}. 

In magma ocean worlds, thick water cloud layers could form \citep[see, e.g.,][]{Lebrun2013,Marcq2017} as the outgassed steam atmospheres cool. These cloud layers could help prevent the loss of planetary infrared radiation to space, and if  thick water clouds were to extend over
the region sampled by transmission spectroscopy, for instance, they might significantly reduce the strength of the spectral features detected. 

\begin{figure}[!hbt]
\hspace{-30pt}
\includegraphics[trim=2cm 0cm 0 7cm ,clip,width=0.56\textwidth]{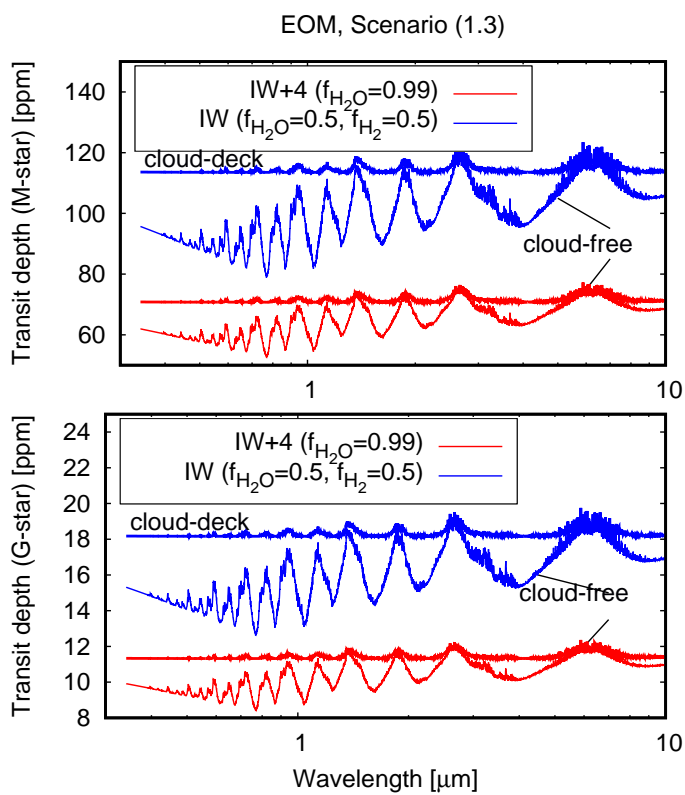}
 \centering
 \caption{Atmospheric transit depth of a cloud-free planetary atmosphere around an M star (top) and G star (bottom) at the EOM phase  for an atmosphere simulated from IW+4 mantle buffer (red) and IW mantle buffer (blue) is compared with the transit depth for cloudy scenarios ($\alpha=0$ and enhanced cross-section as compared to the Rayleigh scattering).}
 \label{clouds}
\end{figure}

In Fig.~\ref{clouds} we show the atmospheric transit depth calculated using Eq.~\eqref{tdepth} for planet Earth orbiting an M-type star (0.4 solar radii) in the top panel and a G-type star (bottom panel) for two different redox states of the mantle, IW+4 and IW. A clear increase in transit depth is visible for the IW buffer with 50\% \ch{H2O} and 0.5\% \ch{H2} because the molecular mass of the atmosphere is lower than in the pure \ch{H2O} case (IW+4 buffer). Results also suggest a substantial increase in the transit depth when observing via an M-type star because its stellar radius is smaller, thus favoring the possibility of detecting atmospheric species. We also compared the cloud-free scenarios with the cloudy simulations. With a cloud deck (see Section~\ref{rad}), the spectral features are weakened (depending upon the cloud cross section), as shown in Figure~\ref{clouds}. The results of the comparison between cloud-free and cloudy scenarios ($\alpha=0$ and enhanced scattering cross section as compared to Rayleigh) agree well with \cite{Turbet2019}.  The transit depths in the optically thick cloudy scenarios can be considered as maximum values (because thick clouds effectively block shortwave radiation) compared with the smaller transit depths of the clear atmosphere cases without clouds. A sensitivity study by \cite{Sarah2018} that involved increasing the haze-scattering cross section also suggested a weakening in spectral features when a global layer of Rayleigh-scattering haze was added to the TRAPPIST-1 d, e, and f planets.  A more involved sensitivity study including the geometry of the cloud, such as its cross section \citep{Kitzmann2011b,Kitzmann2011a}, requires the use of detailed cloud microphysics \citep{Zsom2012} that is applicable for Earth-like exoplanets. We  did not explore this here and aim to investigate it in future related studies.

\subsection{Application to exoplanets}

Several hundred rocky exoplanets classified as super-Earths with mass ranging between $M = 1-8 M_{\rm E}$ and radius $R = 1-1.8 R_{\rm E}$ are being discovered by various current space missions such as TESS\footnote{Transiting Exoplanet Survey Satellite} and will be discovered by future missions such as PLATO\footnote{PLAnetary Transits and Oscillations of stars}, JWST\footnote{James Webb Space Telescope}, and CHEOPS\footnote{CHaracterising ExOPlanets Satellite}. Atmospheric characterization of these planets via spectroscopic analysis can benefit greatly from interior geochemical models that consider outgassing during the magma ocean phase \citep{Bower2019} and volcanic outgassing after the magma ocean phase \citep{Ortenzi2020}.  \cite{Ortenzi2020} showed that the redox state of the mantle affects the atmospheric thickness and overall evolution, thus constraining the interior chemical state of rocky exoplanets. Puffy outgassed atmospheres for hot, molten Earth-like planets often lead to an effect known as ``radius inflation" that depends upon the initial volatile inventory of \ch{H2O} and \ch{CO2} \citep{Bower2019} and on the irradiation that they receive from the star \citep{Turbet2019}. 

Furthermore, as shown in Fig.~\ref{clouds}, the redox state of the mantle and initial inventory of volatiles can induce strong variations in the transit depths for planets orbiting G and M stars. This would translate into changes in the planetary radius to be observed by missions such as TESS and PLATO \citep{Rauer2014}. However, the detectability of molecular spectral features may be severely affected by the presence of clouds \cite[see e.g.][]{Fauchez2019}.

\cite{Kalt2007} modeled the observable spectra of Earth-like planets about two billion years ago to a present-day atmosphere. They reported a number of constraints, such as the resolution of the telescopes for detecting the molecular species in the atmosphere. It is therefore important to characterize the atmosphere of Earth through its geological history and that our study present useful insights on this. Recent interior-atmosphere modeling studies have shown that pure steam-based atmospheres of highly irradiated exoplanets orbiting M dwarfs such as GJ 1214 b \citep{Schaefer2016} and LHS 3844 b are subject to strong XUV of the star and could lose the outgassed atmosphere and become a bare rock \citep{Kreidberg2019}. See also \cite{Kite2020} for a detailed study of atmospheric loss and subsequent revival relevant for exoplanets. In all these models, the water is dissociated into hydrogen and oxygen, which escape to space, and some remaining oxygen is dissolved into the magma ocean or remains in the atmosphere \citep[e.g.,][]{Schaefer2016}.

We have provided important testable predictions to inform future space missions such as the JWST, Ariel\footnote{Atmospheric Remote-sensing Infrared Exoplanet Large Survey} \citep{Tinetti2018,Billy2019}, LIFE\footnote{Large Interferometer for Exoplanets} \citep{Quanz19}, and ELT\footnote{Extremely Large Telescope} \citep{hiRes2013} that will characterize planetary atmospheres in terms of detecting species such as CO, CO$_2$ , and \ch{H2O} in spectral observations of Earth-sized planets.

Planets in close-in orbits around cooler stars are favored targets in exoplanetary science because of their
improved contrast ratios, higher transit frequency, and increased geometric transit probability \cite[][]{Scalo2007,Shields2016}, for example. Furthermore, the detectability of atmospheric spectral signals from MO exoplanets \citep{Ito2015} and thick atmospheres with \ch{H2O} and \ch{CO2} with a high sensitivity and large spectral coverage of the JWST (NIRSpec, MIRI) instruments \citep{Ducrot2020} would bring new possibilities to constrain such atmospheres.


\section{Conclusions}\label{s6}

The main conclusions of this study are listed below.
\begin{enumerate}
    \item 
The redox state of the mantle can affect not only atmospheric composition, but also the vertical temperature structure and therefore the mixing and the hydrological cycle. These interior-atmospheric couplings could be important for steam-condensation timescales and accordingly, for the formation of the  Earth oceans. 
 
\item Reduced atmospheres emit thermal radiation more strongly, which leads to faster cooling than in oxidized atmospheres because the latter feature strong absorbers such as H$_2$O.
 
\item Reduced atmospheres (\ch{CO} and \ch{H2}) have much larger spectral features in transmission than oxidized atmospheres (\ch{H2O} and \ch{CO2}) because the scale heights are larger.

\item Thick optical clouds and hazes can mute or weaken the spectral signatures of molecules in the transmission spectra and enhance the transit depth because of the cloud cover. It therefore becomes difficult to probe cloudy atmospheres. On the other hand, clouds absorb much of the radiation and reduce the outgoing longwave radiation by 1-2 orders of magnitude \citep{Marcq2017}, leading to longer magma ocean cooling timescales.
 
\item  The interplay between outgassing and escape  suggests that outgassing of hydrogen proceeds faster than the atmospheric escape of \ch{H2} and that enough outgassed hydrogen is available to form an atmosphere over the magma ocean duration of $\sim$1 Myr. The timescale for total \ch{H2} mass loss, however, is estimated to be within some dozen million years.  Furthermore, a more sophisticated hydrodynamical code to study the loss of \ch{H2} and the heavy atmospheric species \ch{H2O} and \ch{CO2} is desirable and planned for future studies. 
 
 \end{enumerate}

\bibliographystyle{aa}
\bibliography{references}

\section*{Acknowledgements}
We thank the anonymous reviewer and the editor for their valuable comments that greatly enhanced the quality of the manuscript. N.K. and G.O. acknowledge funding
from the Deutsche Forschungsgemeinschaft (DFG, German Research Foundation) – Project-ID 263649064 – TRR 170. This is TRR 170 Publication No. 117. M.G. acknowledges funding from DFG - Project-ID GO 2610/1-1. F.S. acknowledges DFG project SCHR 1125/3-1. 

\appendix

\section{Gas constant for a mixture}
\label{App1}
When a mixture of two gases $a$ and $b$, with partial  pressure $p_a$ and $p_b$ , exists at the same volume $V$ and temperature $T$, the mixture behaves like a perfect gas, and according to Dalton's law of partial pressures,
\begin{equation}
\label{dl}
(p_a+p_b)V = (m_aR_a + m_b R_b)T.
\end{equation}
Because the mixture behaves like a perfect gas, for $m=nM$ moles of a gas, the total pressure is given by
\begin{equation}
\label{dl1}
PV = m R_{\rm mix}T,
\end{equation}
where $R_{\rm mix}$ is the gas constant for the mixture. From Eqs.~\eqref{dl} and \eqref{dl1}, $mR_{\rm mix} = m_aR_a + m_bR_b$. Hence,
\begin{equation}
R_{\rm mix} = \frac{m_aR_a+m_b R_b}{m}.
\end{equation}
In terms of the molar volume fraction,
\begin{equation}
R_{\rm mix} = \frac{n_aM_aR_a+n_bM_b R_b}{\sum_i n_i M_i}
.\end{equation}
Because $MR = \bar R$, where $\bar R$ is the universal gas constant (= 8.31415 J kg$^{-1}$ K$^{-1}$), this equation can be written as
\begin{equation}
\label{r2}
R_{\rm mix} = \frac{n_a \bar R + n_b \bar R}{n_a M_a + n_b M_b}.
\end{equation}
For a single-component gas $a$, $R = \bar R / M_a$. For a two-component gas mixture, Eq.\eqref{r2} can be written as
\begin{equation}
R_{\rm mix} = \bar R \left(\frac{n_a + n_b}{\mu}\right),
\end{equation}
where $\mu = n_a M_a + n_b M_b$ is the mean molecular weight of the mixture. For an ith-component mixture of gases,
\begin{equation}
\label{rmix}
R_{\rm mix} = \bar R \left(\frac{\sum n_i}{\sum n_i M_i}\right).
\end{equation}

\newpage
\section{Volatile speciation model output}
\label{App2}

\begin{table*}[hbt]
    \centering
    \caption{Cases for BOM used to produce Figure~\ref{OLR}. Cases 1, 4, and 6 are similar to scenarios 1.1, 1.2, and 1.3, respectively, which are also presented in Table~\ref{buffer1}. Columns 2 and 3 show the initial assumed mole fractions arising from \ch{H2O} and \ch{CO2} outgassing, which is the input to the speciation model. The assumed buffer values for strongly reducing (IW-4),  reducing (IW),
 and highly oxidizing (IW+4) have been used to study the effect of speciation under these conditions. The four columns on the right show final outgassed species from the speciation model. For these scenarios, the surface p, T setting is fixed to be $T_{\rm s} = 3300$ K and $P_{\rm s}=76.7$ bar. The $p_{\rm boA}$ calculated from the new molecular weight of the atmosphere is also shown.  The most dominant species in the atmosphere are marked in bold for each of the cases. }

\begin{tabular}{|c||c|c|c|c||c|c|c|c|c||}
 \hline
 Case&\multicolumn{2}{c|}{Initial outgassing} &$P_{\rm s}$ (bar)&Buffer&$p_{\rm boA}$ (bar)&\multicolumn{4}{c|}{Final  outgassing} \\
 & $f^{\rm init}_{\rm \ch{H_2O}}$&$f^{\rm init}_{\rm \ch{CO2}}$&&&&$f_{\rm \ch{CO2}}$& $f_{\rm \ch{H2O}}$&$f_{\rm \ch{H2}}$&$f_{\rm CO}$\\
 \hline
 \hline
1&0.05&0.95 &76.7& IW-4&48&0.0018 &0.0007& 0.049& {\bf 0.94}\\
&&&&IW&53.2&0.15& 0.029& 0.02&{\bf 0.79}\\
& &&&IW+4&75.3&{\bf 0.90}& 0.049& 0.000& 0.046\\
\hline
 2&0.25&0.75  &76.7&IW-4&44.14&0.0014 &0.036 &{\bf 0.24}&{\bf 0.75}\\
&&&&IW&52.83&0.12&{\bf 0.15}& 0.10 &{\bf 0.62}\\
& &&&IW+4&75.43&{\bf 0.71}&{\bf  0.245}& 0.0016& 0.036\\
\hline
3&0.5&0.5  &76.7&IW-4&37.43&0.00 &0.007 &{\bf 0.49}&{\bf 0.50}\\
&&&&IW&52.12&0.08&{\bf 0.30}& 0.20 &{\bf 0.42}\\
& &&&IW+4&75.6&{\bf 0.47}&{\bf  0.29}& 0.003& 0.024\\
\hline
4&0.75&0.25 &76.7&IW-4& 27.1&0.000& 0.01& {\bf 0.74}& {\bf 0.25}\\
&&&&IW&51&0.040& {\bf 0.44}& {\bf 0.30}& {\bf 0.20}\\     
& &&&IW+4&75.8&{\bf 0.24}& {\bf 0.74}& 0.005& 0.0122\\
\hline
5&0.95&0.25  &76.7&IW-4&14.0&0.0 &0.013 &{\bf 0.94}&0.0049\\
&&&&IW&49.6&0.008&{\bf 0.56}& {\bf 0.38} &0.042\\
& &&&IW+4&76.1&0.047&{\bf  0.94}& 0.0063& 0.002\\
\hline
6&1.0&0.00  &76.7&IW-4&9.5&0.0 &0.014 &{\bf 0.98}&0.00\\
&&&&IW&49.1&0.00&{\bf 0.60}& {\bf 0.40} &0.00\\
& &&&IW+4&76.2&0.00&{\bf  0.99}& 0.0069& 0.00\\

\hline
\end{tabular}
\label{case1}
\end{table*}

\begin{table*}[hbt]
    \centering
    \caption{Same as for Table~\ref{case1}, but for EOM  cases  used  to produce Figure~\ref{OLR}. Cases 1, 4, and 6 are similar to scenarios 2.1, 2.2, and 2.3, respectively, which are also presented in Table~\ref{buffer2}. The surface p,T setting is fixed at  $T_{\rm s} = 1650$ K and $P_{\rm s} = 395$ bar. }
    \begin{tabular}{|c||c|c|c|c||c|c|c|c|c||}
 \hline
 Case&\multicolumn{2}{c|}{Initial outgassing } &$P_{\rm s}$ (bar)&Buffer&$p_{\rm boA}$ (bar)&\multicolumn{4}{c|}{Final  outgassing} \\
 & $f^{\rm init}_{\rm \ch{H_2O}}$&$f^{\rm init}_{\rm \ch{CO2}}$&&&&$f_{\rm \ch{CO2}}$& $f_{\rm \ch{H2O}}$&$f_{\rm \ch{H2}}$&$f_{\rm CO}$\\
 \hline
 \hline
1&0.05&0.95 &395&IW-4&247.5& 0.003 &0.0005& 0.049& {\bf 0.94}\\
&&&&IW&283.7&0.22& 0.025& 0.025&{\bf 0.72}\\
& &&&IW+4&390.4&{\bf 0.92}& 0.05& 0.000& 0.03\\
\hline
2&0.25&0.95 &395&IW-4&227.2&0.002 &0.002 &{\bf 0.25}&{\bf 0.74}\\
&&&&IW&277.2&0.176&0.124&  0.125 &{\bf 0.57}\\
& &&&IW+4&390.6&{\bf 0.72}&{\bf  0.24}& 0.002& 0.023\\
\hline
3&0.5&0.5 &395&IW-4&192.4&0.0015 &0.005 &{\bf 0.49}&{\bf 0.49}\\
&&&&IW&266&0.11&{\bf 0.25}& {\bf 0.25} &{\bf 0.38}\\
& &&&IW+4&390.7&{\bf 0.48}&{\bf  0.49}& 0.005& 0.0157\\
\hline
4&0.75&0.25 &395&IW-4& 139&0.0007& 0.007& {\bf 0.74}& {\bf 0.25}\\
&&&&IW&248.9&0.058& {\bf 0.37}& {\bf 0.37}& 0.19\\     
& &&&+4&391&{\bf 0.24}& {\bf 0.74}& 0.007& 0.007\\
\hline
5&0.95&0.25 &395&IW-4&70.66&0.00 &0.009 &{\bf 0.94}&0.05\\
&&&&IW&226.7&0.011&{\bf 0.47}& {\bf 0.47} &0.038\\
& &&&IW+4&391.3&0.048&{\bf  0.94}& 0.0094& 0.001\\
 \hline
6&1.0&0.0 &395&IW-4&47.3&0.00 &0.009 &{\bf 0.99}&0.00\\
&&&&IW&219&0.00&{\bf 0.50}& {\bf 0.50} &0.00\\
& &&&IW+4&391.5&0.00&{\bf  0.99}& 0.0099& 0.00\\
\hline

\end{tabular}

\label{case2}
\end{table*}

\newpage
\section{OLR of \ch{H2} atmospheres}
\label{App3}
\begin{figure}[!h]
\centering
 \includegraphics[width=0.5\textwidth]{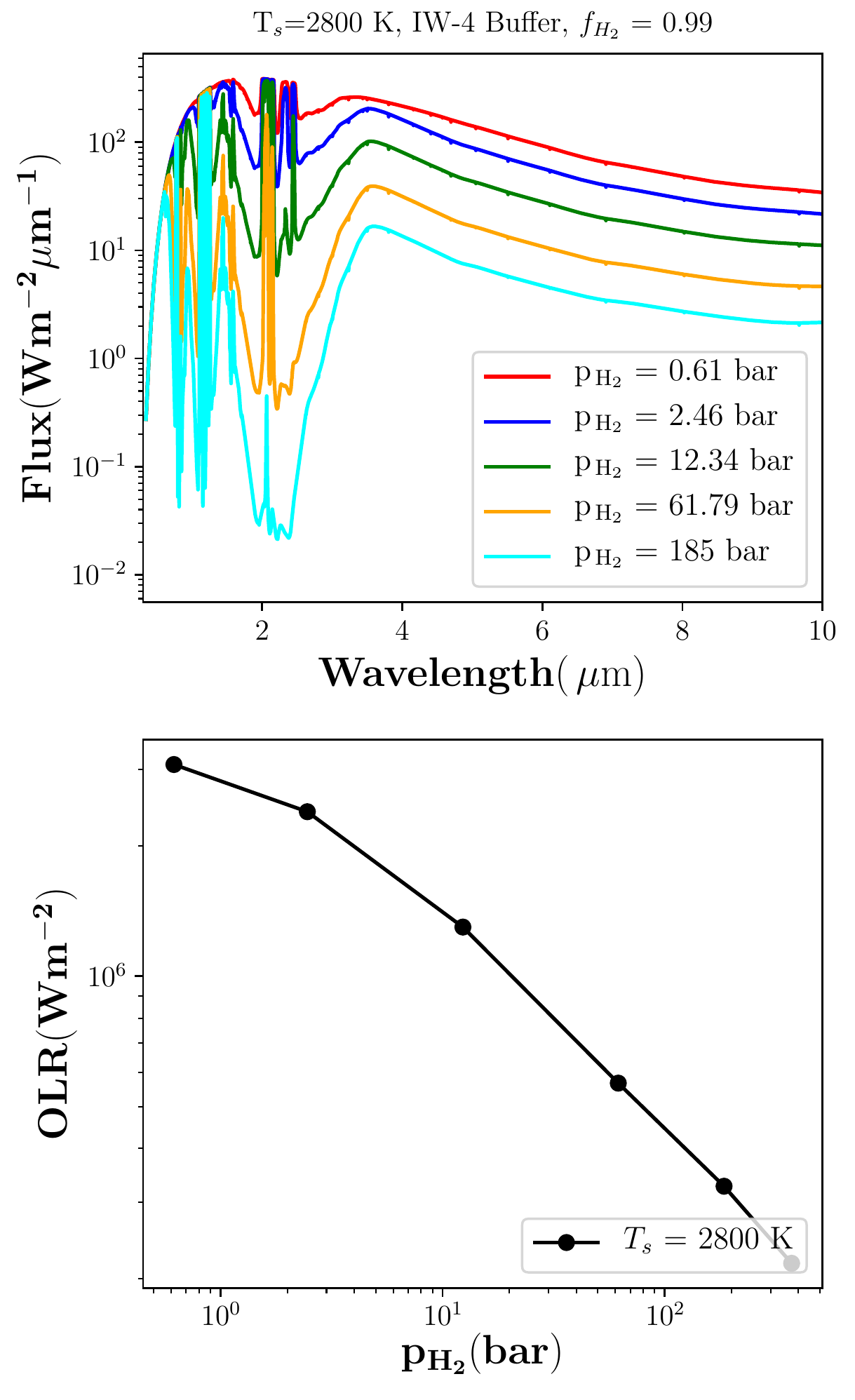}
 \caption{{\emph Top}: Thermal emission spectra for an \ch{H2}-dominated atmosphere overlying a highly reduced mantle (IW-4) for a varying surface partial pressure of hydrogen as shown in the legend.  {\it  Bottom}: OLR plotted against p$_{\ch{H2}}$ for a fixed surface temperature of $T_{\rm s} = 2800$ K showing a decrease in OLR with increase in \ch{H2} surface partial pressure.}
 \label{OLR_pressure}
\end{figure}

\end{document}